\RequirePackage{fix-cm}
\documentclass[smallextended]{svjour3}       % onecolumn (second format)
\smartqed  % flush right qed marks, e.g. at end of proof
\usepackage{graphicx}
 \journalname{myjournal}
\usepackage{dcolumn}
\usepackage{bm}
\usepackage{braket}
\usepackage{amsmath}
\usepackage{amsfonts}
\usepackage{amssymb}
\usepackage{xcolor,soul}
\newcommand{\rmd}{\mathrm{d}}
\newcommand{\aver}[1]{\langle#1\rangle}
\graphicspath{{./figures/}}
\begin{document}

\title{Estimation of entanglement in bipartite systems directly from tomograms%\thanks{Grants or other notes
}


\author{B.~Sharmila\and
        S.~Lakshmibala\and
        V.~Balakrishnan %etc.
}


\institute{B. Sharmila \at
              Department of Physics, Indian Institute of Technology Madras, Chennai 600036, India\\
%              Tel.: +123-45-678910\\
%              Fax: +123-45-678910\\
              \email{sharmilab@physics.iitm.ac.in}           %  \\
           \and
           S. Lakshmibala\at
              Department of Physics, Indian Institute of Technology Madras, Chennai 600036, India\\
%              Tel.: +91-44-22574869\\
%              Fax: +123-45-678910\\
              \email{slbala@physics.iitm.ac.in}           %  \\
           \and
           V. Balakrishnan\at
              Department of Physics, Indian Institute of Technology Madras, Chennai 600036, India\\
              \email{vbalki@physics.iitm.ac.in}
}

\date{Received: date / Accepted: date}

\maketitle

\begin{abstract}
We investigate the advantages of extracting the degree of entanglement in bipartite systems directly from tomograms, as it is the latter that are readily obtained from experiments. This would provide a superior alternative to the standard procedure of assessing the extent of entanglement between subsystems  after employing the machinery of state reconstruction from the tomogram. The latter is both cumbersome and involves  statistical methods, while a direct inference about entanglement from the tomogram circumvents these limitations. In an earlier paper, we had identified a procedure to obtain a bipartite entanglement indicator directly from tomograms. To assess the efficacy of this indicator, we now carry out a detailed investigation using two nonlinear bipartite models by comparing this tomographic indicator with standard markers of entanglement such as the subsystem linear entropy and the subsystem von Neumann entropy and also with a commonly-used indicator obtained from inverse participation ratios. The two model systems selected for this purpose are  a multilevel atom interacting with a radiation field, and a double-well Bose-Einstein condensate. The role played by the specific initial states of these two systems in the performance of the tomographic indicator is also examined. Further, the efficiency of the  tomographic entanglement indicator during the dynamical evolution of the system is assessed from a time-series analysis of the difference between this indicator and the subsystem von Neumann entropy.
\end{abstract}

%Insert your abstract here. Include keywords, PACS and mathematical
%subject classification numbers as needed.

\section{Introduction}
\label{intro}

Quantum state reconstruction seeks to obtain the density matrix and the  corresponding Wigner function from tomograms. This quest  involves statistical procedures which are inherently error-prone.  
It is therefore desirable, as far as possible,  to extract  information about the state (such as quantitative measures of its  nonclassicality and entanglement)  {\em  directly from the tomogram}, avoiding  the reconstruction procedure. This has been demonstrated  in bipartite qubit systems by estimating state fidelity with respect to a specific target state directly from the tomogram~\cite{DSPE1,DSPE2,DSPE3,DSPE4,HuiKhoon}, and comparing the errors that arise with the corresponding errors  in procedures involving complete state reconstruction. 

Obtaining information about a quantum state directly from tomograms for continuous variable processes such as matter-radiation interactions qualifies to be a milestone in state tomography. This is because reconstruction is far more cumbersome in the context of optical tomograms:  the Hilbert space of the field is infinite-dimensional, and homodyne measurements  yield only limited information (namely, expectation values of the density matrix in only a {\em finite} set of  quadrature bases). Even in this case, however, a few results have  been established. It has been shown that it is possible, 
{\em solely} from tomograms, 
 to assess qualitatively whether subsystems are entangled ~\cite{sudhrohithbs}. The squeezing properties 
 of the state 
 and the entanglement between subsystems \cite{pradip2} have been quantified. Further, specific nonclassical effects have been 
 assessed from tomograms during temporal evolution of single-mode and bipartite systems. For instance, qualitative signatures of revivals (when a system returns to its initial state apart from an overall phase) and fractional revivals (when the state of a system is a superposition of 
 `copies' of its initial state) have been identified in tomograms in the case of a single-mode radiation field propagating in a Kerr medium \cite{sudhrohithrev}. Decoherence of entangled bipartite states  has also been investigated using  tomograms \cite{pradip2}. Quadrature and higher-order squeezing of a radiation field subject to cubic nonlinearities have been quantified, and an entanglement indicator  which can be inferred from the tomogram for bipartite systems such as a double-well Bose-Einstein condensate(BEC) has been proposed \cite{sharmila}. 

The last of the foregoing  is an important step in exploiting tomograms, 
 as entanglement is an essential resource in quantum information processing. In this context it is important to consider  interesting phenomena such as sudden death and birth of entanglement ~\cite{eberly},  and its collapse to a constant non-zero value over a significant interval of time~\cite{pradip1},  as have been 
 found  in model systems. Since we are directly concerned with quantum entanglement indicators here, we recall some standard measures of entanglement between the two subsystems $A$ and $B$ of a bipartite system. Two important measures of entanglement  are 
 the subsystem von Neumann entropy 
 (SVNE), given by  $-\mathrm{Tr}\,(\rho_{i} \,\log \,\rho_{i})$ where 
 $\rho_{i}$ ($i=A,B$) is the reduced density matrix of the subsystem, 
  and the subsystem linear entropy (SLE), given by  
  $1-\mathrm{Tr}\,(\rho_{i}^{2})$.  It is evident that  the SVNE and the 
  SLE   involve both off-diagonal and diagonal elements of the density matrix in any given basis. In contrast, the tomogram only provides information about the diagonal elements, although in several complete bases. Even though the tomographic indicator mentioned earlier mimics quite closely the qualitative features of the SVNE and SLE~\cite{sharmila}, it does not qualify to be a measure. It is therefore necessary to carry out a detailed comparative study between the tomographic indicator, on the one hand, and the SLE and SVNE, on the other hand, in order to assess its efficacy and limitations 
  vis-\`a-vis  standard entanglement measures. 

Another interesting entanglement measure with which the tomographic indicator is to be compared  is the inverse participation ratio. In a bipartite system,  this ratio is a measure of the spread of the wave function of the system over subsystem basis states. The ratio itself is defined in terms of the fourth power  of the system  wave function. A procedure to extract the extent of entanglement in spin systems  from the inverse participation ratio has been outlined in \cite{ViolaBrown}, and a 
measure of entanglement $\xi_{\textsc{ipr}}$ has been defined. Extensive examination of the efficacy of $\xi_{\textsc{ipr}}$ in interacting spin systems has been carried out (see, for instance, \cite{HamCorBSSV,IPRent1,IPRent2,IPRent3,IPRent4,IPRent5,IPRent6,IPRent7,IPRent8}).  The  limitations of $\xi_{\textsc{ipr}}$ in capturing the salient features of the SVNE in entangled qubit  systems have been 
pointed out in ~\cite{IPRent2}. We identify the analog of $\xi_{\textsc{ipr}}$ in the 
case of a continuous variable system, and assess the performance of the tomographic indicator relative  to that of  $\xi_{\textsc{ipr}}$.

The models we consider for our purposes describe  two experimentally viable bipartite systems, namely, the double-well BEC with nonlinear interactions between the condensate atoms~\cite{sanz}, and a multilevel nonlinear atomic medium  interacting with a radiation field~\cite{agarwalpuri}. 
Extensive literature on entanglement dynamics exists in the case of these systems. For instance, in a binary Bose-Einstein condensate, the dynamics of the variance of appropriate observables was found to mimic entanglement dynamics~\cite{mistakidis}. Further, the experimental realization of atomic homodyne measurements \cite{BECtomoent}  has enabled quantum state tomography in BECs, while optical homodyne measurements are an integral part of field tomography \cite{QSTopt}.
We have also obtained a long data set of the difference between 
the SVNE and the tomographic indicator, and carried out a time-series analysis  of this difference for several initial states and for 
different amounts of nonlinearity. The ergodicity properties of this difference may be expected to shed light on  
the dynamical behaviour of the tomographic indicator. 

The plan of the rest of this paper is as follows.  In Sec.  
\ref{sec:2} , we describe how the tomographic indicator is obtained,  and relate it to the  participation ratio. In Sec. \ref{sec:models}, we introduce the two bipartite models mentioned above and compare various indicators during dynamical evolution. Section \ref{sec:NTSA} is devoted to the time-series analysis referred to above. 

\section{Bipartite entanglement indicator from the tomogram}
\label{sec:2}
We begin with a quick summary of  the salient features of a tomogram, 
followed by a description of  the procedure for extracting an indicator of bipartite entanglement  solely from the tomogram. 

\subsection{Salient features of a tomogram}
The starting point is a quorum of observables (which can, in principle, be measured through appropriate experiments) whose statistics gives us {\em   tomographically}  complete information about the state. In the case of  a single-mode radiation field, for instance, a quorum is constituted by 
the set of rotated quadrature operators  \cite{VogelRisken,ibort} 
 \begin{equation}
\label{eqn:quadop}
\mathbb{X}_{\theta} = (a^\dagger \,e^{i \theta} + 
a \,e^{-i \theta})/\sqrt{2}
\end{equation}
where $0 \leq \theta <  \pi$,  and 
 $a$ and $a^{\dagger}$ are photon annihilation and creation operators  
 satisfying $[a,a^{\dagger}]=1$. The expectation value of the density matrix $\rho$ can be computed in each complete basis set $\lbrace\ket{X_{\theta},\theta}\rbrace$ for a given $\theta$. The tomogram \cite{VogelRisken,LvovskyRaymer} $w(X_\theta, \theta) = \bra{X_\theta, \theta} \rho \ket{X_\theta, \theta}$ is usually represented as a 
 three-dimensional plot of $w$ versus $X_{\theta}$ (on the $x$-axis) and $\theta$ (on the $y$-axis).

Of immediate relevance to us is the optical tomogram~\cite{ibort} corresponding to a bipartite system with subsystems $A$ and $B$ with rotated quadrature operators
\begin{equation}
\mathbb{X}_{\theta_{A}} = (a \,e^{-i \theta_{A}} + a^{\dagger} \,e^{i \theta_{A}})/\sqrt{2} 
\label{eqn:rot_quad_1}
\end{equation}
and
\begin{equation}
\mathbb{X}_{\theta_{B}} = (b e^{-i \theta_{B}} + b^{\dagger} e^{i \theta_{B}})
/\sqrt{2}.
\label{eqn:rot_quad_2}
\end{equation}
Here $a, a^{\dagger}$ (resp.,  $b, b^{\dagger}$) are the annihilation and creation operators for subsystem $A$ (resp., $B$), 
 and $0 \leq \theta_{A}, \theta_{B} < \pi$. The bipartite tomogram is given by 
\begin{align}
\nonumber w(X_{\theta_{A}},\theta_{A}; & X_{\theta_{B}},\theta_{B}) =\\
& \bra{X_{\theta_{A}},\theta_{A};X_{\theta_{B}},\theta_{B}} \rho_{AB} \ket{X_{\theta_{A}},\theta_{A};X_{\theta_{B}},\theta_{B}}, 
\label{eqn:tomo_defn}
\end{align}
where $\rho_{AB}$ denotes the bipartite density matrix. 
This is just a straightforward extension of the single-mode tomogram. 
Here, $\mathbb{X}_{\theta_{i}}\ket{X_{\theta_{i}},\theta_{i}}=X_{\theta_{i}}\ket{X_{\theta_{i}},\theta_{i}}$ ($i=A,B$),  and $\ket{X_{\theta_{A}},\theta_{A};X_{\theta_{B}},\theta_{B}}$ stands for 
 $\ket{X_{\theta_{A}},\theta_{A}} \otimes \ket{X_{\theta_{B}},\theta_{B}}$.  The normalization condition is
\begin{equation}
\int_{-\infty}^{\infty} \rmd X_{\theta_{A}} \int_{-\infty}^{\infty} \rmd X_{\theta_{B}} w(X_{\theta_{A}},\theta_{A};X_{\theta_{B}},\theta_{B}) = 1
\label{eqn:tomo_norm}
\end{equation}
for every value of $\theta_{A}$ and $\theta_{B}$. The \textit{reduced} tomogram
 for subsystem $A$ is 
\begin{align}
\nonumber w_{A}(X_{\theta_{A}},\theta_{A}) & = \int_{-\infty}^{\infty} \rmd X_{\theta_{B}} w(X_{\theta_{A}},\theta_{A};X_{\theta_{B}},\theta_{B})\\[4pt]
& = \bra{X_{\theta_{A}},\theta_{A}} \rho_{A} \ket{X_{\theta_{A}},\theta_{A}},
\label{eqn:red_tomo_1}
\end{align}
where $\rho_{A}=\mathrm{Tr_{B}}(\rho_{AB})$ is the reduced density matrix 
of the subsystem $A$.  A similar definition holds for the subsystem $B$.

The bipartite tomographic entropy and the subsystem tomographic entropy 
are  important concepts  that we require for our tomographic  entanglement indicator. The former is given by 
\begin{align}
S(\theta_{A},\theta_{B}) = 
- & \int_{-\infty}^{\infty} \!\rmd X_{\theta_{A}} \int_{-\infty}^{\infty} \!\rmd X_{\theta_{B}} w(X_{\theta_{A}},\theta_{A};X_{\theta_{B}},
\theta_{B})\times \nonumber\\[4pt] 
& \log \,[w(X_{\theta_{A}},\theta_{A};X_{\theta_{B}},\theta_{B})].
\label{eqn:2_mode_entropy}
\end{align}
The subsystem tomographic entropy is
\begin{align}
S(\theta_{i}) = - &\int_{-\infty}^{\infty} \!\rmd X_{\theta_{i}} w_{i}(X_{\theta_{i}},\theta_{i}) \times \nonumber \\[4pt]
&\log \,[w_{i}(X_{\theta_{i}},\theta_{i})] \;\; (i=A,B).
\label{eqn:1_mode_entropy}
\end{align}
As shown in \cite{sharmila}, the mutual information is given by
\begin{equation}
S(\theta_{A}:\theta_{B}) 
= S(\theta_{A},\theta_{B}) - S(\theta_{A}) - S(\theta_{B}).
\label{mutinfo}
\end{equation}  
The  tomographic entanglement indicator, 
given by 
\begin{equation}
\xi_{\textsc{tei}} = \langle S(\theta_{A}:\theta_{B}) \rangle, 
\label{xiteidefn}
\end{equation} 
is obtained by averaging the mutual information  
over an ideally very large  number of values of  
$\theta_{A}$ and  $\theta_{B}$ over the interval $[0,\pi)$. In practice  
(as shown explicitly \cite{sharmila} in the case of the double-well BEC system), however,  even as few
as $25$ values of $S(\theta_{A}:\theta_{B})$ for $\theta_{A}$ and $\theta_{B}$ chosen at equally-spaced intervals in the range $[0,\pi)$ 
suffice to yield  a $\xi_{\textsc{tei}}$ that 
compares well with  standard entanglement measures such as the SVNE.

\subsection{Participation ratio and entanglement indicator}
The generalized eigenstates of conjugate pairs of quadrature operators constitute a pair of mutually unbiased bases~\cite{xpMUB}, as
\begin{equation}
\left\vert\aver{X_{\theta},\theta|X'_{\theta+\pi/2},\theta+\pi/2}
\right\vert = 1/\sqrt{2 \pi \hbar} \,> 0.
\label{eqn:xpMUB}
\end{equation}
The specific averaging procedure mentioned above obviously involves calculating $S(\theta_{A}:\theta_{B})$ in several sets of mutually unbiased bases. Parallels can be drawn between $\xi_{\textsc{ipr}}$ and $\xi_{\textsc{tei}}$,  as a similar averaging procedure (in this case, over inverse participation ratios) is followed in calculating  $\xi_{\textsc{ipr}}$ \cite{ViolaBrown}.
This is seen by writing the inverse participation ratio  $\eta_{AB}$ corresponding to a given bipartite  pure state 
$\ket{\psi_{AB}}$ in terms of basis states of relevance to the problem at hand, namely, the rotated quadrature basis 
$\lbrace\ket{X_{\theta_{A}},\theta_{A};X_{\theta_{B}},\theta_{B}}\rbrace$, for  specific   values of $\theta_{A}$ and  $\theta_{B}$.  We have 
\begin{equation}
\eta_{AB}= \int_{-\infty}^{\infty} \!\rmd X_{\theta_{A}} \int_{-\infty}^{\infty} \!\rmd X_{\theta_{B}} \big\vert\bra{X_{\theta_{A}},\theta_{A};X_{\theta_{B}},\theta_{B}}\psi_{AB}\rangle\big\vert^{4}.
\end{equation}
 For ease of notation, the  dependence of $\eta_{AB}$ on $\theta_{A}$ and $\theta_{B}$ has been omitted in the left-hand side of the 
 forgoing expression. $\eta_{AB}$ is readily expressed  in terms of the tomogram as 
\begin{equation}
\eta_{AB} = \int_{-\infty}^{\infty} \!\rmd X_{\theta_{A}} \int_{-\infty}^{\infty} \!\rmd X_{\theta_{B}} [w(X_{\theta_{A}},\theta_{A};X_{\theta_{B}},\theta_{B})]^{2}.
\label{eqn:IPR12_defn}
\end{equation}
The inverse participation ratio for each  subsystem is  given by 
\begin{equation}
\eta_{i} = \int_{-\infty}^{\infty} \!\rmd X_{\theta_{i}} [w_{i}(X_{\theta_{i}},\theta_{i})]^{2} \;\; (i=A,B).
\label{eqn:IPRi_defn}
\end{equation}
Note that $\eta_{i}$ depends on $\theta_{i}$. The numerical results reported in the next section seem to suggest  that $(\eta_{A}+\eta_{B}-\eta_{AB})$ averaged over mutually unbiased bases, is equal to ($1-\xi_{\textsc{ipr}}$), although we have not proved this rigorously. 

It is appropriate to compare $\xi_{\textsc{tei}}$  and $\xi_{\textsc{ipr}}$, 
because both these quantities   
can be obtained directly from the tomogram, and the similarity in form of the two relations  $(1-\xi_{\textsc{ipr}})=\aver{\eta_{A}+\eta_{B}-\eta_{AB}}$  and $\xi_{\textsc{tei}} =\aver{ S(\theta_{A},\theta_{B}) - S(\theta_{A}) - S(\theta_{B})}$ is manifest. (Here $\aver{\cdot\cdot}$ denotes the relevant 
 average in each case.) 
 In obtaining $\xi_{\textsc{tei}}$, it is evident that (in practice) a good approximation would be  to average \textit{only} over the dominant values of $S(\theta_{A}:\theta_{B})$. We have compared the results obtained numerically by averaging over the complete set of values of $S(\theta_{A}:\theta_{B})$ and those obtained by averaging only over the dominant values (values that exceed the mean value by one standard deviation). We find that the results do not change significantly in the qualitative features. Hence in the following sections, we present results retaining only the dominant values as it is computationally efficient. 
 We denote the entanglement indicator thus obtained by  $\xi'_{\textsc{tei}}$, and in subsequent sections compare $\xi'_{\textsc{tei}}$ with the SLE and $\xi_{\textsc{ipr}}$.

\section{Entanglement indicators in generic bipartite models}
\label{sec:models} We investigate the properties of $\xi'_{\textsc{tei}}$  in two generic bipartite models describing,  
respectively,   a multilevel atom interacting with a radiation field \cite{agarwalpuri} and the double-well BEC \cite{sanz}. Both systems
are inherently nonlinear and  provide an ideal platform to examine nonclassical effects as the corresponding state evolves in time. 
In this paper, we consider initial states of the 
total bipartite system that are pure states subsequently governed by unitary evolution. This ensures that the SVNE (and similarly, the SLE) remain 
independent of the subsystem label ($A$ or $B$) for all $t$.

\subsection{Atom-field interaction model}

We consider a nonlinear multilevel atomic medium coupled with strength $g$ to a radiation field of frequency $\omega_{F}$. The effective Hamiltonian (setting $\hbar=1$) is \cite{agarwalpuri}  
\begin{equation}
H_{\textsc{AF}}=\omega_{F} a^{\dagger} a + \omega_{A} b^{\dagger} b + \gamma b^{\dagger \,2} b^{2} + g ( a^{\dagger} b + a b^{\dagger}).
\label{eqn:HAF}
\end{equation}
$a, a^{\dagger}$ are photon  annihilation and creation operators. 
The multilevel atom is modelled as an 
oscillator with harmonic frequency 
$\omega_{A}$ and  ladder operators $b, b^{\dagger}$.
The nonlinear atomic medium is  effectively described by the Kerr-like term in $H_{\textsc{AF}}$ with strength $\gamma$.  A  variety of initial states of the total system has been judiciously selected in order to explore the  range of possible nonclassical effects during time evolution.  
The {\em unentangled}  initial states considered correspond to the atom in its ground state $\ket{0}$ and the field in either a coherent state (CS), 
 or an $m$-photon added coherent state ($m$-PACS). In the photon 
number basis the CS $\ket{\alpha}$ ($\alpha \in \mathbb{C}$) is 
of course  given by 
\begin{equation}
\ket{\alpha} = 
 e^{-|\alpha|^{2}/2} 
\sum_{k=0}^{\infty} \frac{\alpha^{k}}{\sqrt{k!}} \ket{k}.
\label{csdefn}
\end{equation} 
The normalized $m$-PACS  $\ket{\alpha,m}$ 
(where $m$ is a positive integer), 
which possesses a precisely  quantified departure from 
perfect coherence, is 
given by 
\begin{equation} 
\ket{\alpha,m} = 
\frac{a^{\dagger\,m}\ket{\alpha}}
{\langle\alpha
\vert  a^{m}a^{\dagger\,m}\ket{\alpha}}
= \frac{a^{\dagger\,m}\ket{\alpha}}
{\sqrt{m!\,L_{m}(-|\alpha|^{2})}}\,,
\label{mpacsdefn}
\end{equation}
where $L_{m}$ is the Laguerre polynomial of order $m$. 
We also consider 
two {\em entangled} initial states, namely, the binomial state 
$\ket{\psi_{\rm{bin}}}$ and the two-mode squeezed state $\ket{\zeta}$. 
These states are defined as follows. The total number operator 
\begin{equation}
N_{\rm{tot}}= a^{\dagger}a + b^{\dagger}b
\label{Ntot}
\end{equation}
 commutes with 
$H_{\textsc{AF}}$, and $\ket{\psi_{\rm{bin}}}$ is an eigenstate of this 
operator with eigenvalue $N$ (a non-negative integer). In 
explicit form, it is given by 
\begin{equation}
\ket{\psi_{\rm{bin}}} = 2^{-N/2} \sum_{n=0}^{N} 
{\tbinom{N}{n}}^{1/2}
 \ket{N-n;n},
\label{eqn:BS_defn}
\end{equation}
where $\ket{N-n;n} \equiv  \ket{N-n}\otimes\ket{n}$, 
the product state corresponding to 
the field and the atom  
 in the respective 
 number states $\ket{N-n}$ and $\ket{n}$. 
 The two-mode squeezed state is given by 
\begin{equation}
\ket{\zeta} = e^{\zeta^{*} a b - \zeta a^{\dagger} b^{\dagger}} \ket{0;0},
\label{eqn:sq_state}
\end{equation} 
where $\zeta \in \mathbb{C}$ and  $\ket{0;0}$ is the 
 product state corresponding to 
 $N = 0, n = 0$. 
 
\begin{figure}[h]
\centering
\includegraphics[width=0.4\textwidth]{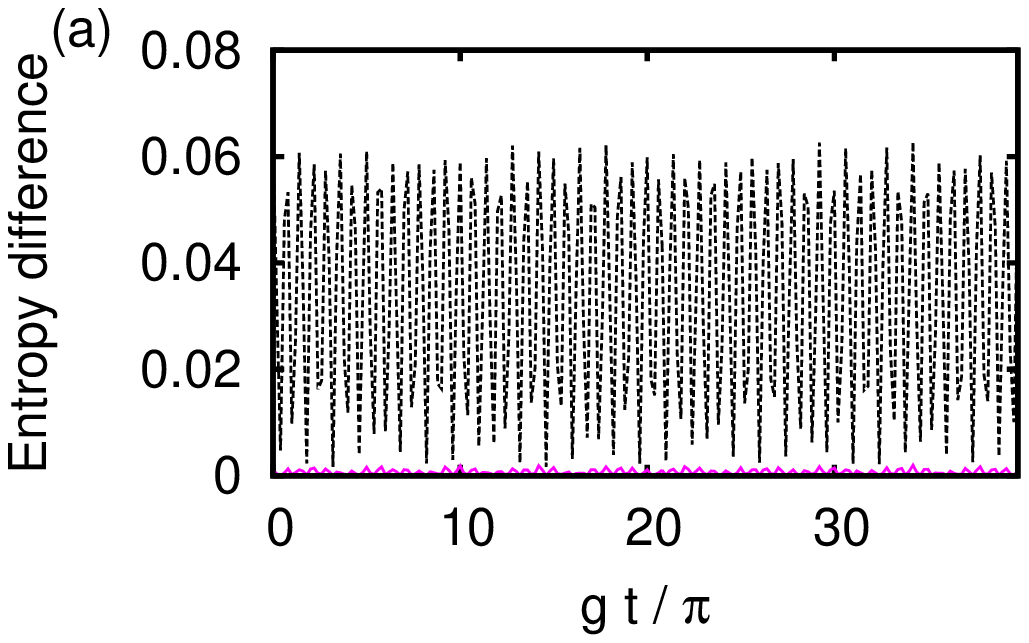}
\includegraphics[width=0.4\textwidth]{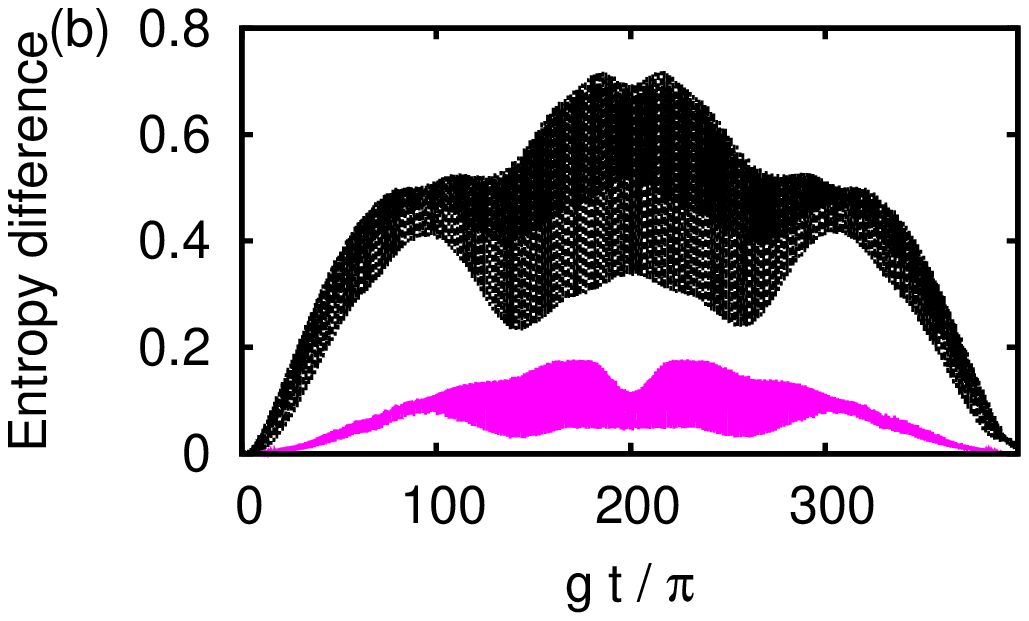}
\caption{$d_{1}(t)$ (black dashed curve) and $d_{2}(t)$ (pink curve) versus scaled time $g t/ \pi$ with $\omega_{F}=1, \omega_{A}=1, \gamma=1$. (a) $g=0.2$, initial squeezed state $\ket{\zeta}, \,\zeta = 0.1$ (b) $g=100$, 
initial state $\ket{\alpha} \otimes \ket{0},  |\alpha|^{2}=1$. }
\label{fig:SLE_vs_SVNE}
\end{figure}

\begin{figure}[h]
\centering
\includegraphics[width=0.4\textwidth]{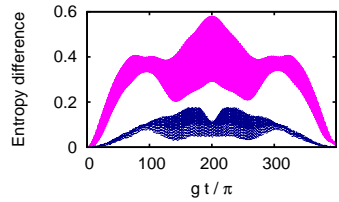}
\caption{$d_{2}(t)$ (blue dotted curve) and $\Delta(t)$ (pink curve)
versus  time,  with $\omega_{F}=1$, $\omega_{A}=1$, $\gamma=1$, $g=100$,  initial state $\ket{\alpha} \otimes \ket{0}$, $|\alpha|^{2}=1$.}
\label{fig:comp_diff_SVNE}
\end{figure}

\begin{figure}[h]
\centering
\includegraphics[width=0.4\textwidth]{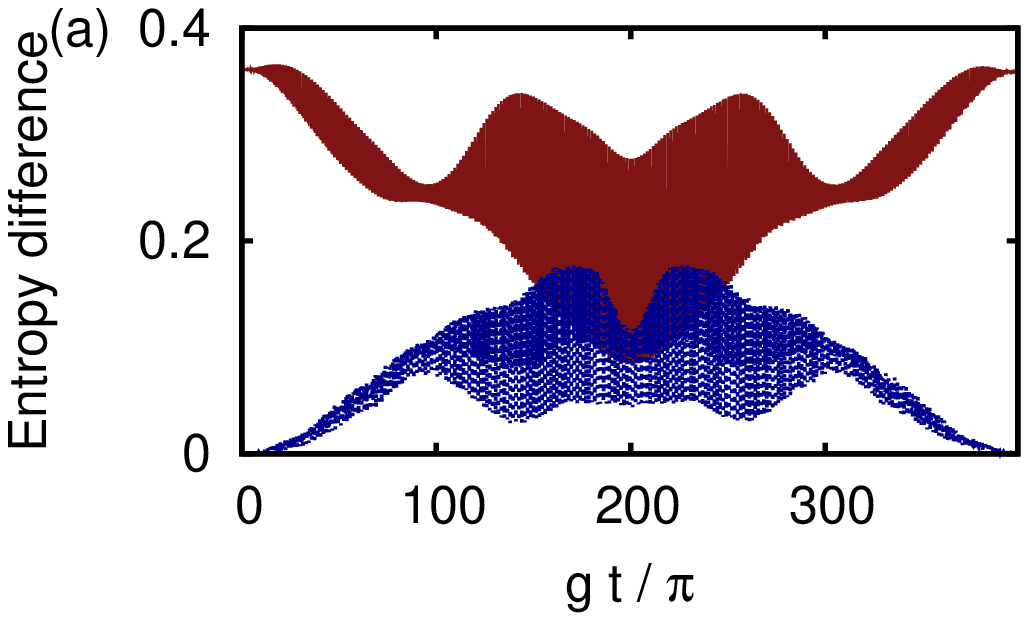}
\includegraphics[width=0.4\textwidth]{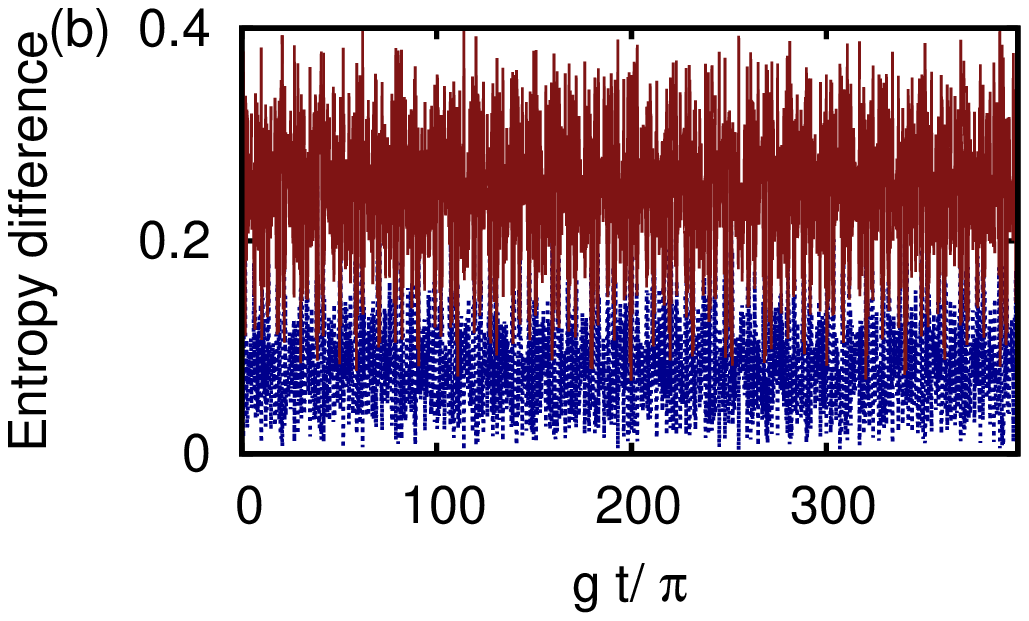}
\includegraphics[width=0.4\textwidth]{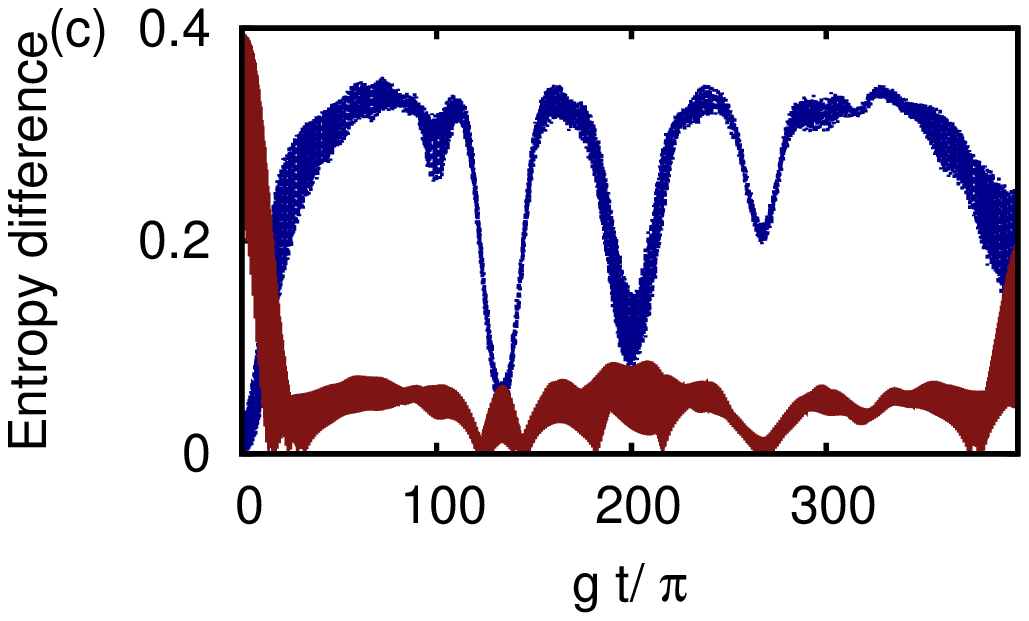}
\includegraphics[width=0.4\textwidth]{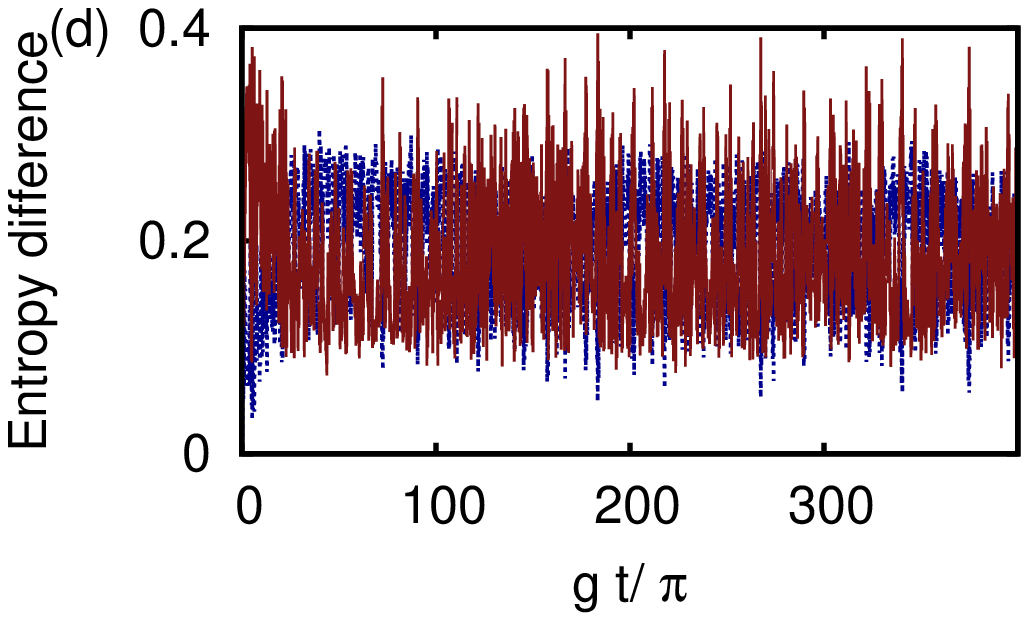}
\caption{$d_{2}(t)$ (blue dotted curve) and $d_{3}(t)$ (brown curve) 
versus time,  with $\omega_{F}=1,  \omega_{A}=1,  \gamma=1$. 
(a) $g=100$, initial state $\ket{\alpha} \otimes \ket{0}$  (b) $g=0.2$, initial state $\ket{\alpha} \otimes \ket{0}$  (c) $g=100$,  
initial state $\ket{\alpha,5} \otimes \ket{0}$ (d) $g=0.2$,   
initial state $\ket{\alpha,5} \otimes \ket{0}$. 
$|\alpha|^{2} = 1$ in all cases.}
\label{fig:comp_diff_1}
\end{figure}

Corresponding to these initial states we have numerically generated tomograms at approximately 2000 instants, separated by a time step 0.2 $\pi / g$. From these,  we have obtained $\xi'_{\textsc{tei}}$ and  the differences
\begin{equation}
d_{1}(t) = \vert {\text{SVNE}} - \xi'_{\textsc{tei}} \vert, \;\;
d_{2}(t) = \vert {\text{SLE}} - \xi'_{\textsc{tei}} \vert
\label{d1d2defns}
\end{equation}
 as the system evolves. These differences are plotted against 
the scaled time $g t / \pi$ in Fig. \ref{fig:SLE_vs_SVNE}(a) 
for an initial two-mode squeezed state, and  in
Fig. \ref{fig:SLE_vs_SVNE}(b) for a factored product of a  CS and atomic ground state $\ket{0}$. From these plots it is evident that $\xi'_{\textsc{tei}}$ is in much better agreement with the SLE than with the SVNE over the time interval considered, independent of the parameter values and the nature of the initial state.  We therefore choose the SLE as the reference entanglement indicator.
Next, we compare $d_{2}(t)$ with the difference 
\begin{equation}
\Delta(t) = \vert {\text{SVNE}} - {\text{SLE}} \vert.
\label{Deltadefn}
\end{equation} 
We have verified that, In all the cases considered, 
$\Delta(t) >  d_{2}(t)$ (see, for instance, Fig. \ref{fig:comp_diff_SVNE}). 
In what follows,  we therefore focus only on   $d_{2}(t)$ and the difference 
\begin{equation}
d_{3}(t) = \vert {\text{SLE}} - \xi_{\textsc{ipr}}\vert.
\label{d3defn}
\end{equation}
This comparison brings out interesting features of both the indicators. When the strength of the nonlinearity is low relative to that of the 
coupling (e.g., $\gamma/g=0.01$), it is known \cite{sudh_rev_moments} 
that  full and fractional revivals occur, and  
entanglement measures may be  expected to display signatures of 
these revival phenomena.
Figure \ref{fig:comp_diff_1} (a) shows that,  at the 
revival time  $g T_{\rm rev}/\pi=400$, $\xi'_{\textsc{tei}}$ agrees with 
the SLE much more closely  than $\xi_{\textsc{ipr}}$ does. Further, over the entire time interval $(0, T_{\rm rev})$, $d_{2}(t)$ is significantly smaller than $d_{3}(t)$. This feature holds even for larger values of the ratio $\gamma/g$, as may be seen in 
Fig. \ref{fig:comp_diff_1} (b).   
$\xi'_{\textsc{tei}}$ is therefore to be favored over $\xi_{\textsc{ipr}}$ as an entanglement indicator. The time evolution 
 of the difference $d_{2}(t)$ is drastically different from 
 that of  $d_{3}(t)$ for initial field states that depart from ideal coherence. In this case, over the entire time considered, $\xi_{\textsc{ipr}}$ performs significantly better than $\xi'_{\textsc{tei}}$ for small values of $\gamma/g$, as in Fig. \ref{fig:comp_diff_1} (c). As  the value of $\gamma/g$ 
 is increased, the two indicators have essentially the same  behavior, 
 as shown in Fig. \ref{fig:comp_diff_1} (d).
%\noindent

\begin{figure}[h]
\centering
\includegraphics[width=0.4\textwidth]{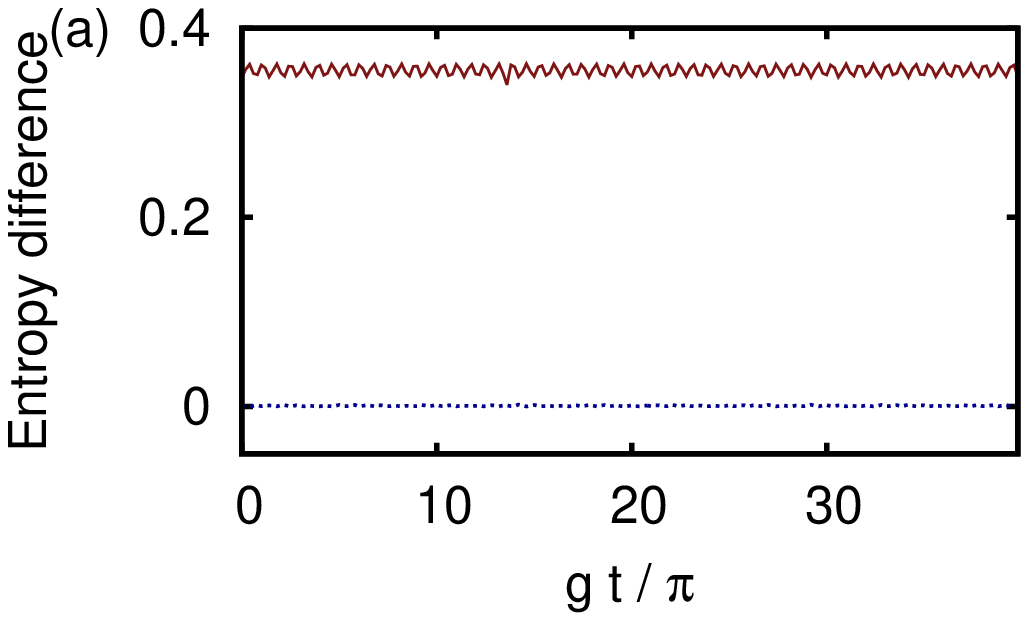}
\includegraphics[width=0.4\textwidth]{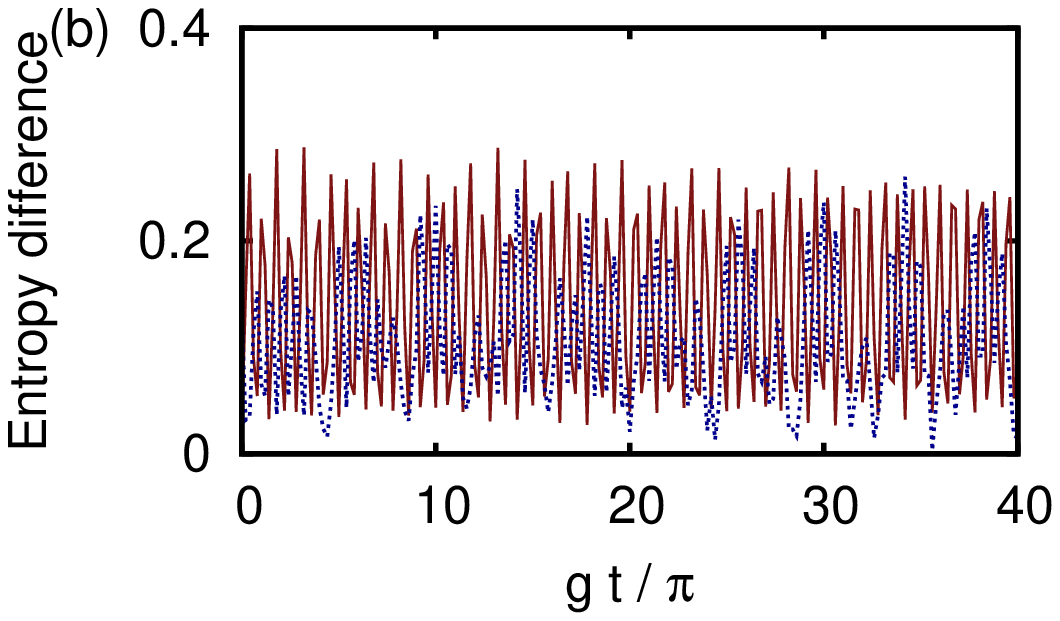}
\caption{$d_{2}(t)$ (blue dotted curve) and $d_{3}(t)$ (brown curve) 
versus time, with $\omega_{F}=1,  \omega_{A}=1,  
\gamma=1, g=0.2$, for the  initial two-mode squeezed state 
$\ket{\zeta}$.  (a) $\zeta=0.1$ (b) $\zeta=0.7$. }
\label{fig:comp_diff_3}
\end{figure}

We turn now to  entangled initial states. In the case of the two-mode squeezed state $\ket{\zeta}$, we see from Figs. \ref{fig:comp_diff_3} (a) and (b) that  $\xi'_{\textsc{tei}}$ fares much better than 
$\xi_{\textsc{ipr}}$ over the entire time interval  considered,  for small values of  $\zeta$.   With an increase in the value of $\zeta$,  both the indicators show comparable departures from SLE.

\begin{figure}[h]
\centering
\includegraphics[width=0.4\textwidth]{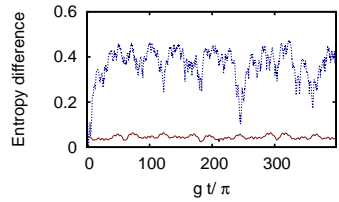}
\caption{$d_{2}(t)$ (blue dotted curve) and $d_{3}(t)$ (brown curve) 
versus time,  with $\omega_{F}=1,  \omega_{A}=1, \gamma=1, 
g=0.2$. Initial state $\ket{\psi_{\rm{bin}}}$ with $N=10$.}
\label{fig:comp_diff_4}
\end{figure}

In the case of   an initial binomial state, on the 
other hand,   $\xi_{\textsc{ipr}}$ fares significantly better 
than  $\xi'_{\textsc{tei}}$. This can be understood by examining the Hamming distance between  the 
basis states constituting  the binomial state. We define this distance in the context of  continuous variables by extrapolating the idea of Hamming distance for qudits given below. The Hamming distance between two qubits  $\ket{u}$  and $\ket{v}$ ($u,v=0,1$) is $1$ if $u \neq v$,  and $0$ if $u=v$. It is evident that  this can be extended to qudits in a straightforward fashion. 
The generalized Hamming distance (see, for instance,  
\cite{ViolaBrown}),  i.e., the distance between two bipartite qudit states $\ket{u_{1};u_{2}}$ (where 
$u_{1}, u_{2}=0,1,\dotsc,  d-1$) and $\ket{v_{1};v_{2}}$ (where 
$v_{1}, v_{2}=0,1,\dotsc, d-1$), is $0$ if $u_{1}=v_{1}$ and $u_{2}=v_{2}$. The Hamming distance is $1$ if $u_{1} = v_{1}$ and 
$u_{2}\neq v_{2}$, or vice versa. The distance is $2$ (the 
maximum possible value in bipartite systems) if 
$u_{1} \neq v_{1}$ and $u_{2} \neq v_{2}$.   Further, the efficacy of 
$\xi_{\textsc{ipr}}$ 
as an entanglement indicator increases with an 
increase in the Hamming distance.  This indicator is 
therefore especially useful  for states which are Hamming-uncorrelated (i.e., separated by a Hamming distance equal to 2) \cite{ViolaBrown,HamCorBSSV}.

In the case of interest to us, both the subsystems are 
infinite-dimensional. We would like  to examine 
whether the efficacy of  $\xi_{\textsc{ipr}}$ is correlated with 
the Hamming distance in this case as well.  
For this purpose, we extend the 
notion of the Hamming distance between two 
unentangled basis states in a straightforward manner:
The distance between $\ket{m;n}$ and $\ket{p;q}$ (where 
$m,n,p,q = 0, 1, 2, \dotsc$ {\it ad   inf.}) is  equal to its maximum value  of $2$ if $m\neq p$ and $n\neq q$; 
$1$ if   $m = p, n \neq q$ or vice versa;  and $0$ if 
$m = p, n = q$. 
 Note that the binomial state  
 $\ket{\psi_{\rm{bin}}}$ 
 can be expanded as a superposition of states which are Hamming-uncorrelated (Eq. (\ref{eqn:BS_defn})). 
Figure \ref{fig:comp_diff_4} shows that in this case, too, 
 $\xi_{\textsc{ipr}}$ is a significantly better entanglement indicator 
than $\xi'_{\textsc{tei}}$.

\subsection{The double-well BEC model}

The effective Hamiltonian for the system 
(setting $\hbar = 1$) is given by \cite{sanz}
\begin{equation}
H_{\textsc{BEC}}
=\omega_{0} N_{\rm{tot}} + \omega_{1} (a^{\dagger} a - b^{\dagger} b) + U  N_{\rm{tot}}^{2} - \lambda (a^{\dagger} b + a b^{\dagger}).
\label{eqn:HBEC}
\end{equation}
$N_{\rm{tot}} = a^{\dagger} a + b^{\dagger} b$ as before, 
but $(a,a^{\dagger})$ and $(b,b^{\dagger})$ are now 
 the  boson annihilation and creation operators of the atoms in wells $A$ and $B$ respectively. $U$ is the strength of the nonlinearity 
 (both in the individual modes as well as in their interaction),  
 $\lambda$ is the linear interaction strength, and $\omega_{0}, \omega_{1}$ are constants. As in the previous instance,  we select a representative variety of initial states: 
 (i) the unentangled direct product $ \ket{\alpha_{a}, m_{1}} 
 \otimes \ket{\alpha_{b}, m_{2}}$ of boson-added 
 coherent states of atoms in the wells $A$ and $B$ respectively, 
 where $\alpha_{a},\alpha_{b} \in \mathbb{C}$; (ii)  
 the binomial state  $\ket{\psi_{\rm{bin}}}$ (Eq. (\ref{eqn:BS_defn})), and (iii) the two-mode squeezed vacuum state $\ket{\zeta}$ (Eq. (\ref{eqn:sq_state})),  with the understanding that the basis states are now product states of the species in the two wells. 
 
In each of these cases, we must first obtain  
the state of the system at 
any time $t \geq 0$ as it evolves under the Hamiltonian 
$H_{\textsc{BEC}}$. It turns out that, in the case of an initial 
state of type (i) above, the state of the system can be 
calculated  explicitly  as a function of $t$, as outlined in the Appendix. In the cases (ii) and (iii), the state vector at 
time $t$ is computed numerically. 
 Using the  results obtained for the state of the system at 
time $t$, we have generated, for each of the initial states 
listed above, tomograms at approximately 1000 instants, separated by a time step 0.001 $\pi/U$.  We have verified that, as in the case of the atom-field 
interaction model, $\xi'_{\textsc{tei}}$ agrees better with the 
SLE than with the SVNE, and that the difference between $\xi'_{\textsc{tei}}$ and the SLE is smaller than that between 
the SVNE and the SLE. In what follows, therefore, we have 
chosen the SLE as the reference entanglement measure and compared  $d_{2}(t)$ with $d_{3}(t)$.

\begin{figure}[h]
\includegraphics[width=0.32\textwidth]{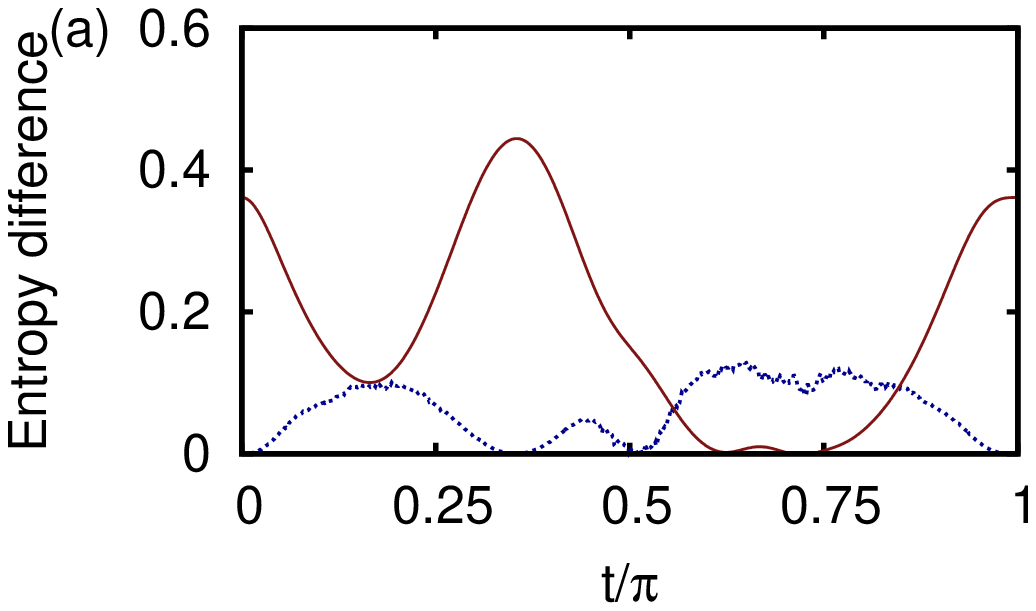}
\includegraphics[width=0.32\textwidth]{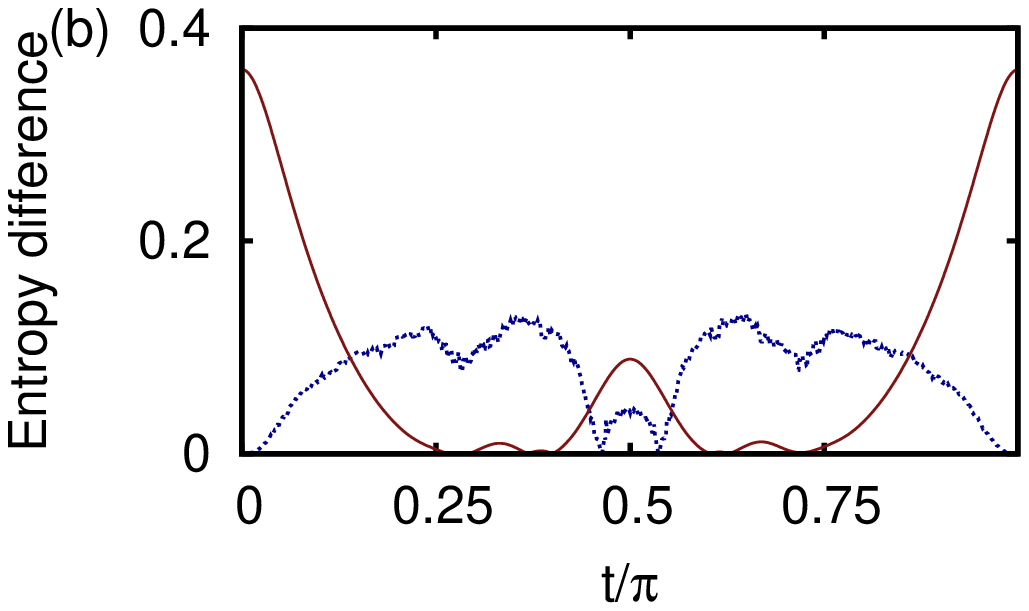}
\includegraphics[width=0.32\textwidth]{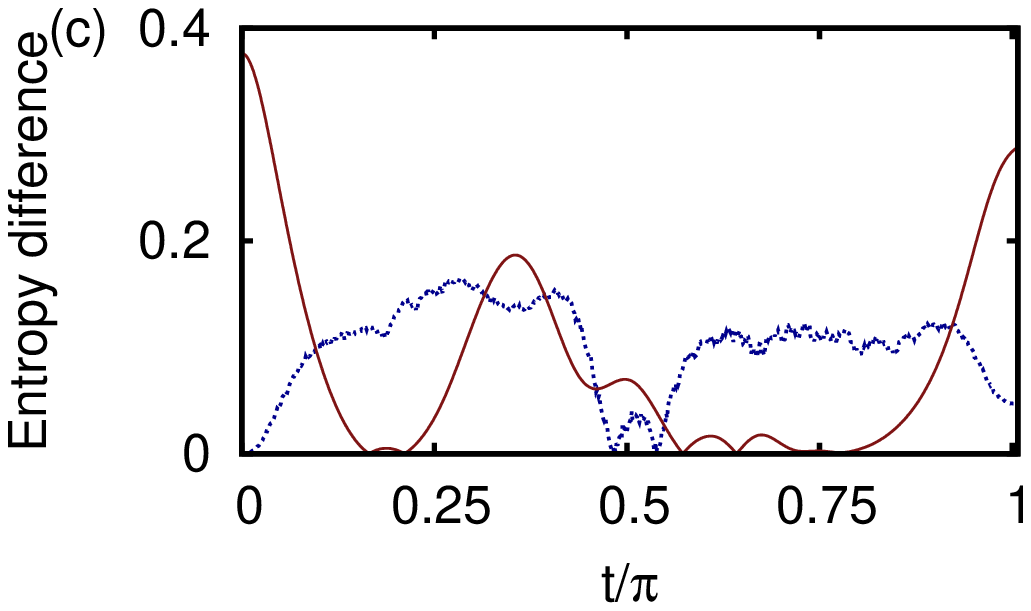}
\caption{$d_{2}(t)$ (blue dotted curve) and $d_{3}(t)$ (brown curve) versus time, with  $\omega_{0}=1, U=1$. (a) $\omega_{1}=1, \lambda=1$, initial state $\ket{\alpha} \otimes \ket{\alpha}$; (b) $\omega_{1}=0.1, \lambda=0.1$,  initial state $\ket{\alpha} \otimes \ket{\alpha}$; (c) $\omega_{1}=1, \lambda=1$,  
initial state $\ket{\alpha,1} \otimes \ket{\alpha}$. 
In all cases, $\vert\alpha\vert^{2} = 1$.}
\label{fig:BEC_comp_diff_1}
\end{figure}

The effective 
frequency parameter for the 
linear part of the Hamiltonian  $H_{\textsc{BEC}}$
is given (see the Appendix) by $\lambda_{1} = (\omega_{1}^{2}+ \lambda^{2})^{1/2}$, 
while the strength of the nonlinearity  is parametrized by $U$. 
Hence,  the relevant ratio for the characterization 
 of the dynamics is $U/\lambda_{1}$. 
Figure  \ref{fig:BEC_comp_diff_1} 
 depicts a representative example of the  temporal behavior of 
 $d_{2}(t)$ and $d_{3}(t)$.  
 The effect of increasing 
  $U/\lambda_{1}$ can be seen by comparing 
 Figs. \ref{fig:BEC_comp_diff_1} 
 (a) and (b), while that of  a departure from coherence of the initial state can be seen by comparing  (a) and  (c). 
We have also carried out analogous studies in the case of 
entangled initial states $\ket{\psi_{\rm{bin}}}$ and $\ket{\zeta}$. 
These results are not presented here owing to limitations of space, 
but the general trends in the behavior of  the entanglement 
indicators in these cases are consistent with, and corroborate, those 
found in the atom-field interaction model. 

\section{Time-series analysis of $d_{1}(t)$}
\label{sec:NTSA}

Finally, we turn to an  assessment of 
 the {\em long-time}  behavior of the tomographic entanglement 
 indicator, by means of a detailed time-series analysis. 

As we have shown in the foregoing, the deviation of   
$\xi'_{\textsc{tei}}$ from the SVNE is much more pronounced than its deviation from the SLE.  It is therefore appropriate to investigate how an initial  difference  between $\xi'_{\textsc{tei}}$ and the SVNE changes with time. 
With this in mind, a time series of $d_{1}(t)$  has been obtained 
for each of  the models at hand, and used to compute local Lyapunov exponents  along the lines customary \cite{AbarbanelBook,NTSA_GP,abarbanel1992} in the study of dynamical systems. This involves  reconstruction of the effective phase space,  estimation of the minimum embedding dimension $d_{\rm{emb}}$ of this space, and  calculation of the exponents themselves. The procedure used is outlined below.

The time series had 20000 data points. The effective phase space was reconstructed using the TISEAN package \cite{tisean}. $100$ different initial values $d_{1}(0)$ were randomly chosen in this phase space. The maximum local Lyapunov exponent corresponding to each $d_{1}(0)$ was computed over the same time interval $L$. (The term `local' refers to the fact that $L$ is much smaller than the time interval over which the maximum Lyapunov exponent $\Lambda_{\infty}$ is obtained in the standard method). The average value $\Lambda_{L}$ of these $100$  maximum local Lyapunov exponents was obtained following the prescription in \cite{abarbanel1992}.   
This procedure was repeated for as many as 14 different values of $L$.  Further, in each case it was verified that,  with an increase in $L$, $\Lambda_{L}$ tends to $\Lambda_{\infty} + (m/L^{q})$, where 
$m$ and $q$ are constants  \cite{abarbanel1992}. Note that two neighbouring initial values of the dynamical variable of interest (in our case $d_{1}(t)$) diverges exponentially with the exponent $\Lambda_{L}$ in $L$ steps. We also present the power spectra of the time series for completeness. We would expect a broadband spectra for chaotic data. On the other hand, a spiky power spectra can point to a possible quasi-periodic behaviour. 
The results are presented below. 
 
\subsection{Atom-field interaction model}

\begin{figure}[h]
\includegraphics[width=0.4\textwidth]{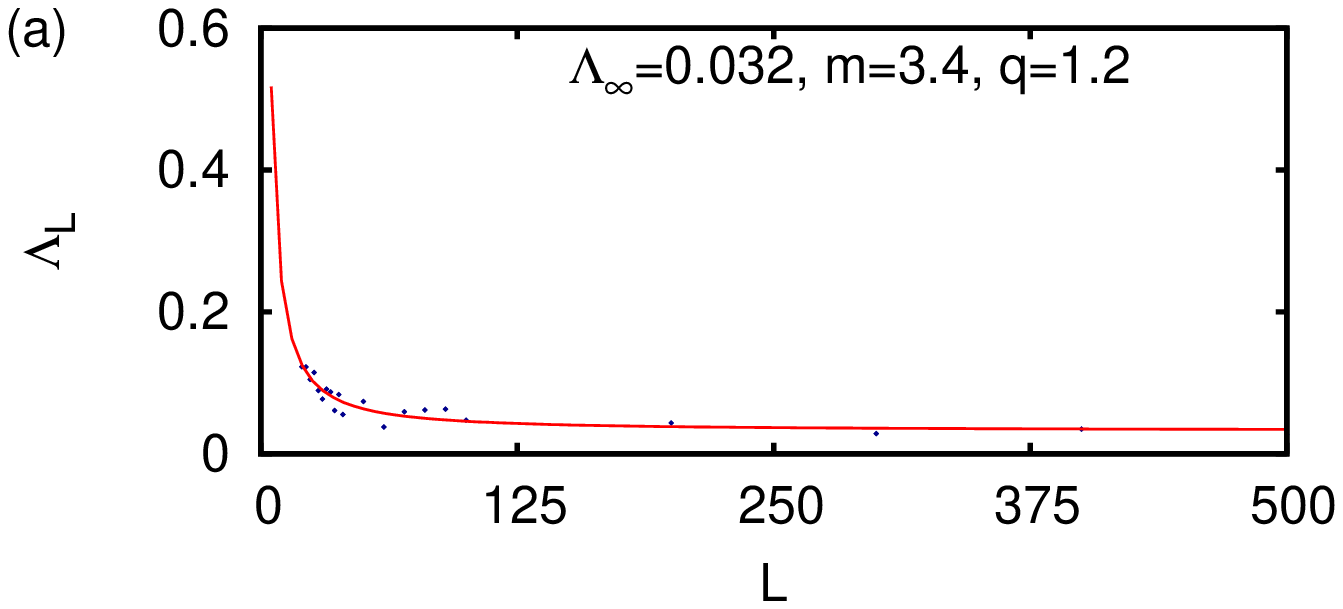}
\includegraphics[width=0.4\textwidth]{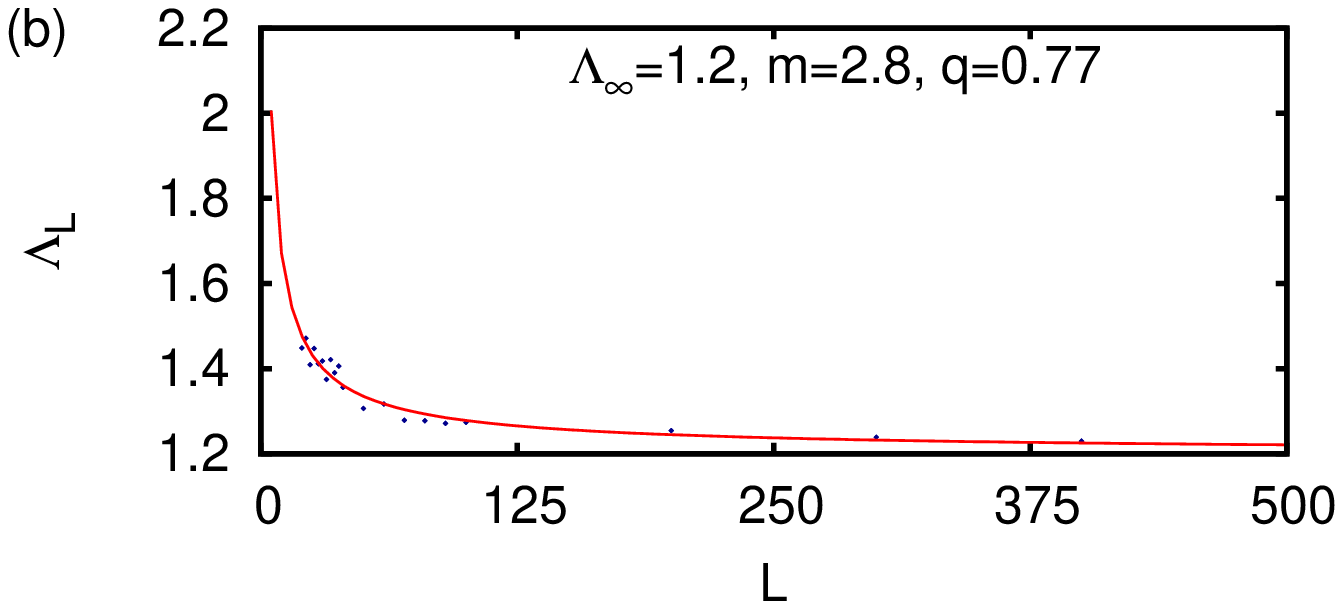}
\includegraphics[width=0.4\textwidth]{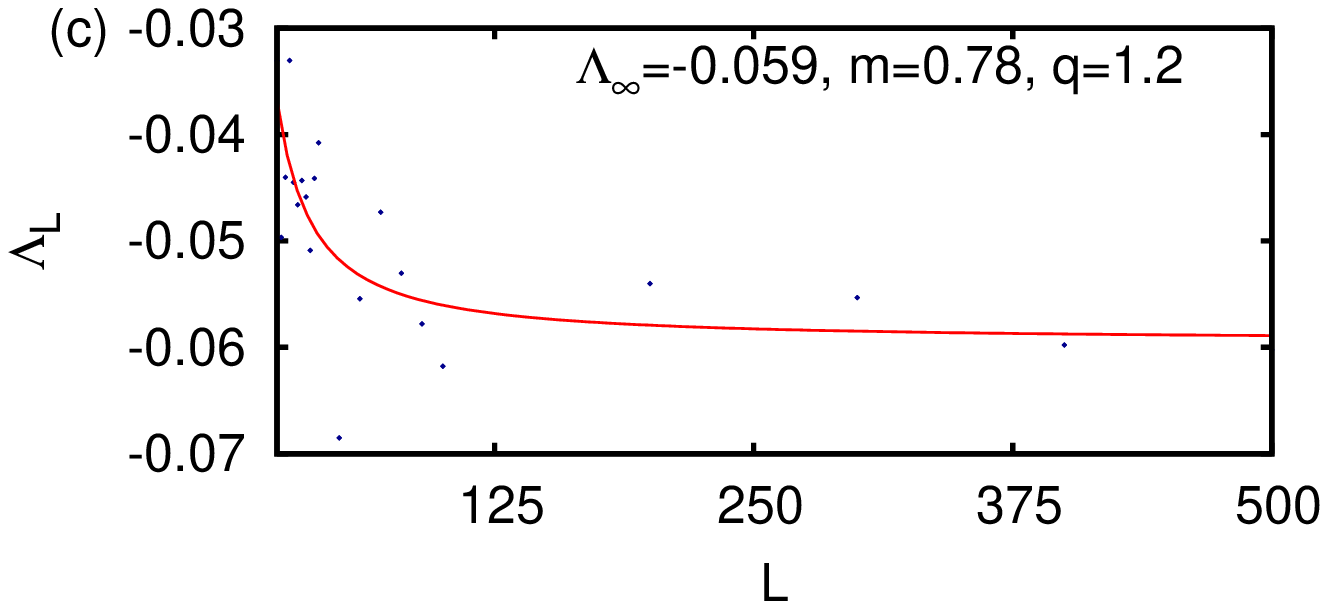}
\includegraphics[width=0.4\textwidth]{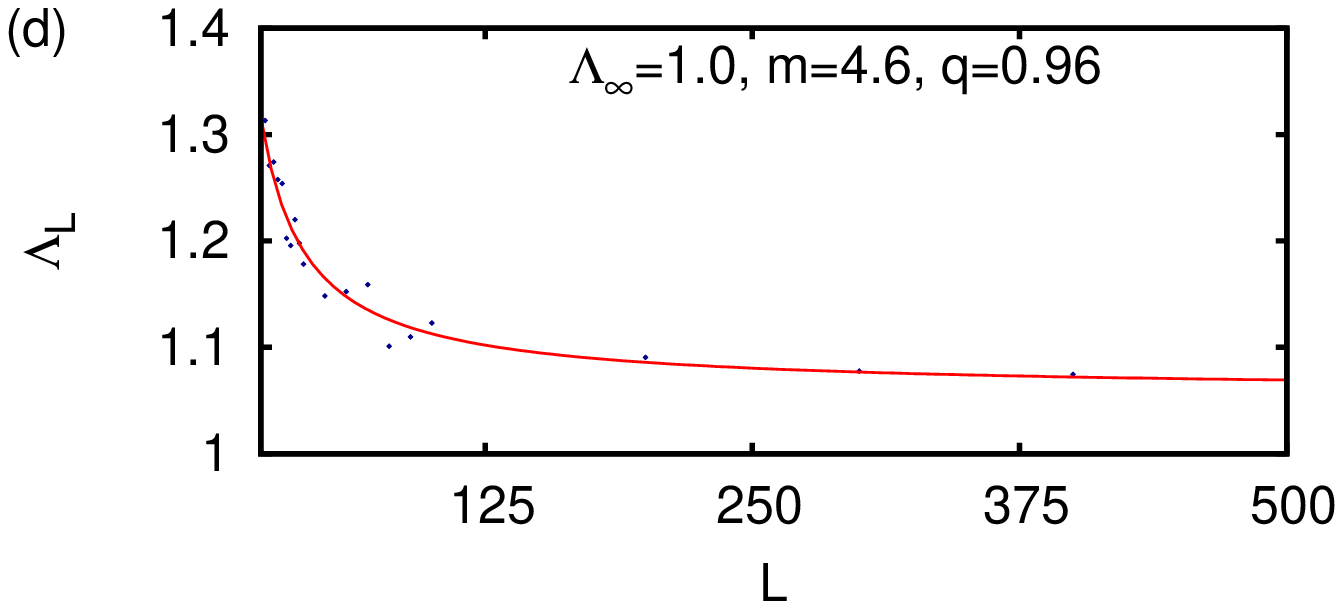}
\caption{$\Lambda_{L}$ obtained from the time series 
of $d_{1}(t)$ (blue crosses) and the fit $\Lambda_{\infty} + (m/L^{q})$ (red curve)  versus $L$,  with $\omega_{F}, \omega_{A}$ and $\gamma$ equal to 1.  Initial state $\ket{\alpha}\otimes \ket{0}$: (a) $g=100$,  $|\alpha|^{2}=1$ (b) $g=100$,  $|\alpha|^{2}=5$ (c) $g=0.2$,  $|\alpha|^{2}=1$. (d) Initial state 
$\ket{\alpha,5}\otimes \ket{0}, g=100, |\alpha|^{2}=1$.}
\label{fig:lambda_L_agarwal}
\end{figure}
The difference $d_{1}(t)$ has been obtained at each instant with time-step $\delta t = 0.1$ for $20000$ time-steps, and the effective phase space has been reconstructed. We see that for both the  initial states $\ket{\alpha} \otimes \ket{0}$ and  $\ket{\alpha,5}\otimes \ket{0}$ with $|\alpha^{2}|=1$ and weak nonlinearity ($\gamma/g=0.01$), $\Lambda_{L}$ is positive, and  
both $\Lambda_{L}$ and  $\Lambda_{\infty}$ are 
larger  for the second initial state (compare Figs. \ref{fig:lambda_L_agarwal} (a) and (d)). $\Lambda_{\infty}$ increases with an increase in $|\alpha|^{2}$ for the 
initial state $\ket{\alpha} \otimes \ket{0}$ (compare Figs. \ref{fig:lambda_L_agarwal} (a) and (b)).  In contrast, for strong nonlinearity (e.g., as in Fig. \ref{fig:lambda_L_agarwal} (c),  
$\gamma/g=5$), $\Lambda_{L}$ is negative.

\begin{figure}[h]
\includegraphics[width=0.4\textwidth]{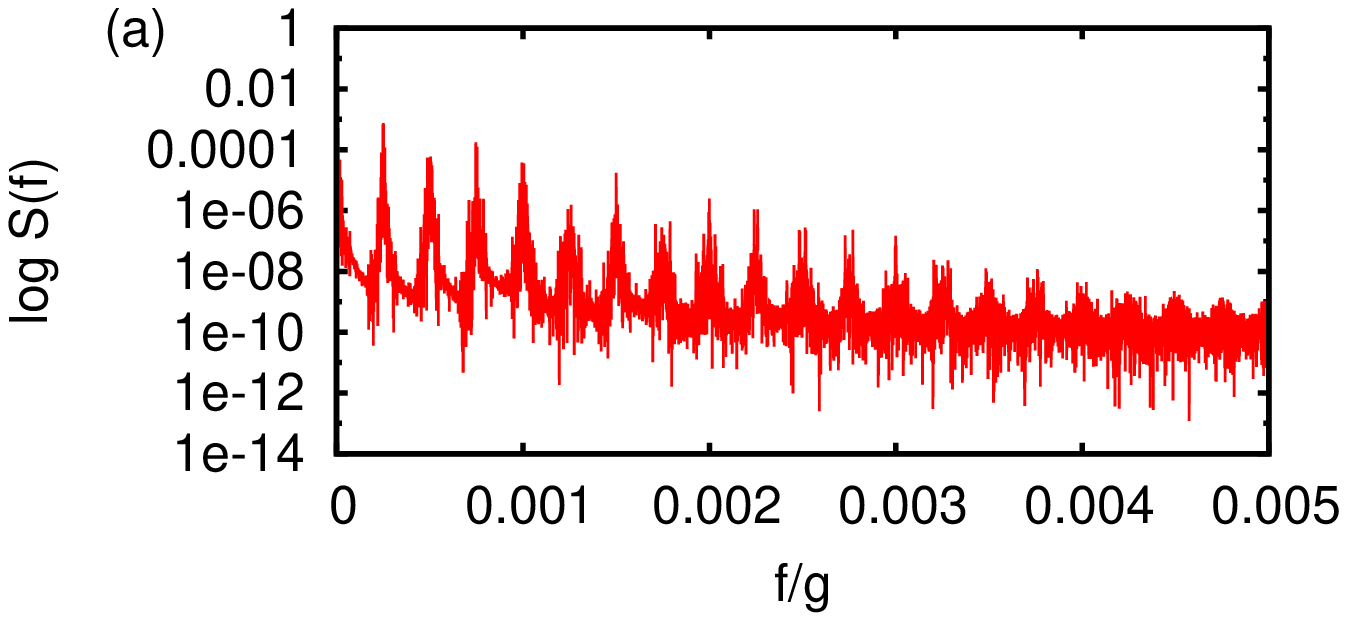}
\includegraphics[width=0.4\textwidth]{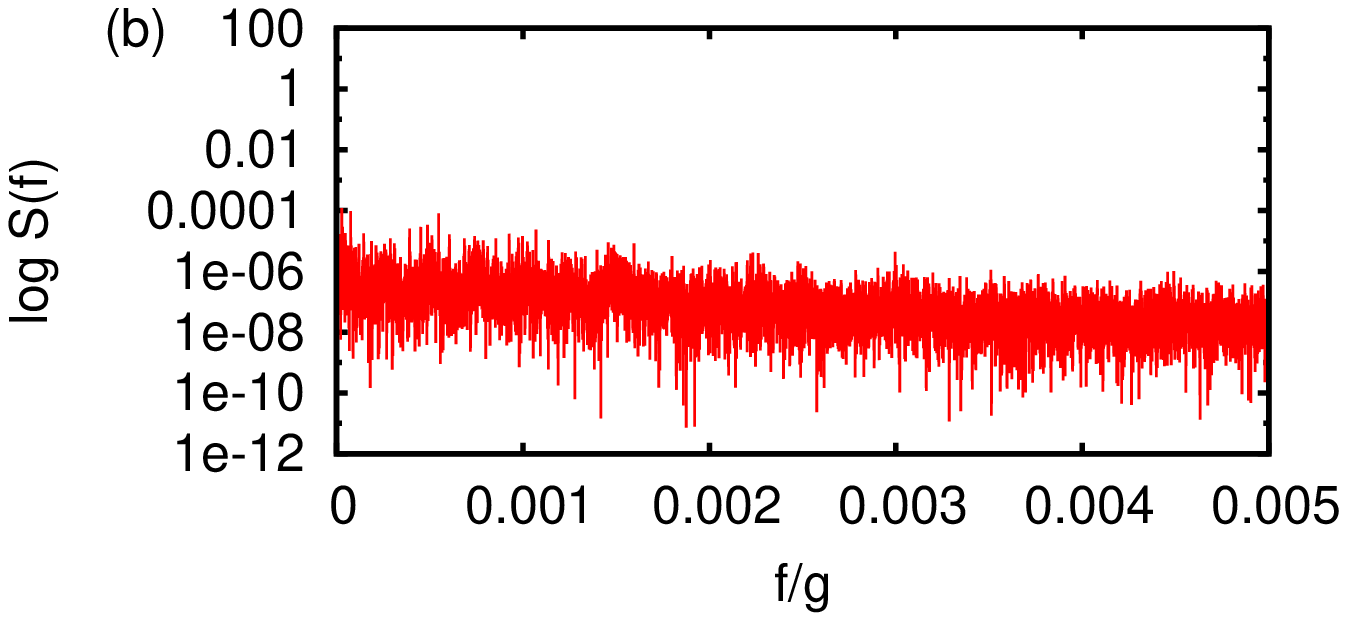}
\includegraphics[width=0.4\textwidth]{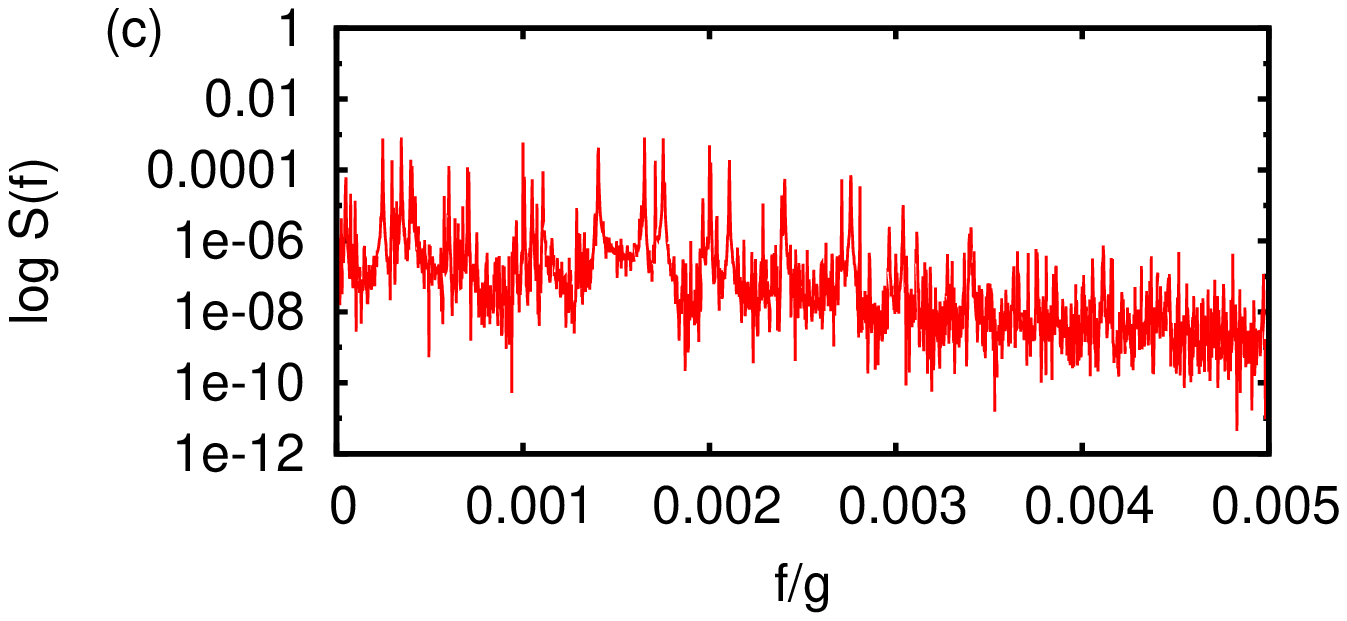}
\includegraphics[width=0.4\textwidth]{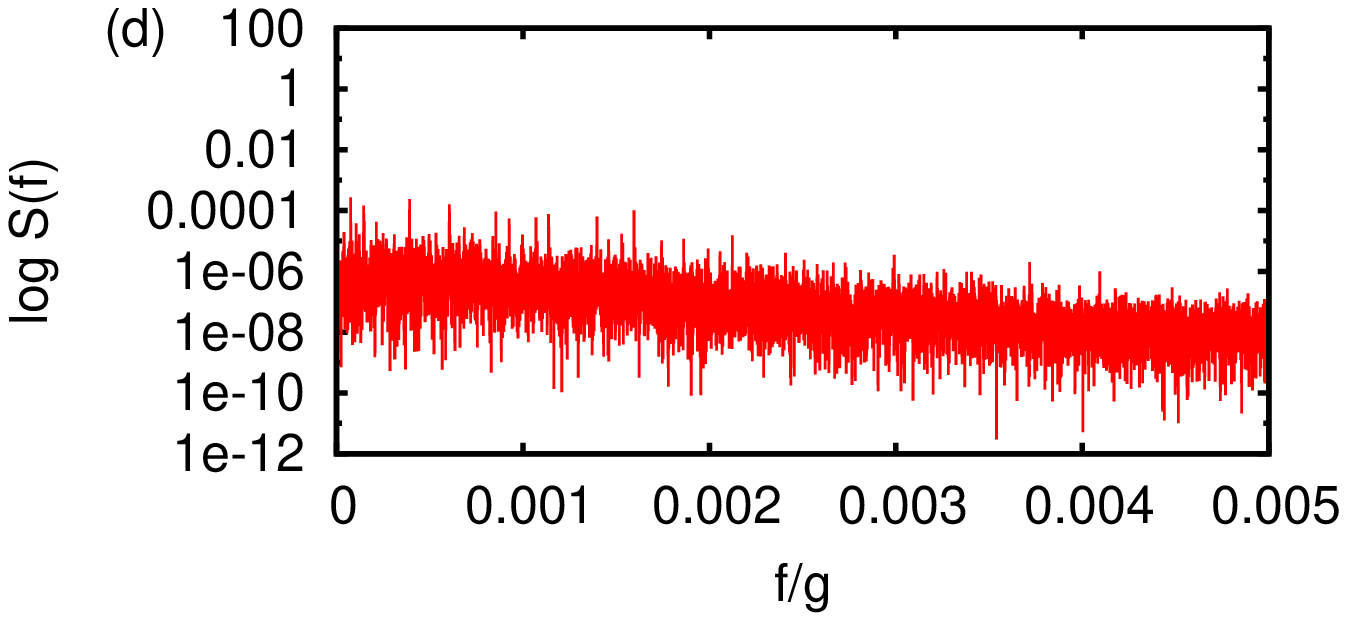}
\caption{Power spectrum $S(f)$ on 
a logarithmic  scale (red curve) versus  $f/g$ for the 
same respective sets of parameters and initial states 
as in Figs. \ref{fig:lambda_L_agarwal} (a)--(d).}
\label{fig:powerspec_agarwal}
\end{figure}

For completeness, we present the power spectrum $S(f)$
of the time series as a function 
of the frequency $f$ in units of $g$, for each of the 
cases corresponding to Figs. 
\ref{fig:lambda_L_agarwal} (a) to (d). 
The nearly quasi-harmonic power spectrum for weak nonlinearity 
(Fig. \ref{fig:lambda_L_agarwal} (a)) 
changes into a broadband 
spectrum with increasing $|\alpha|^{2}$  
(Fig. \ref{fig:lambda_L_agarwal} (b)), while it  loses 
its quasi-harmonicity  without becoming a broadband 
spectrum with increasing nonlinearity
(Fig. \ref{fig:lambda_L_agarwal} (c)).  The lack of 
 coherence in the initial state makes the power spectrum 
broadbanded (Fig. \ref{fig:lambda_L_agarwal} (d)).  

\subsection{The double-well BEC model}

\begin{figure}[h]
\centering
\includegraphics[width=0.4\textwidth]{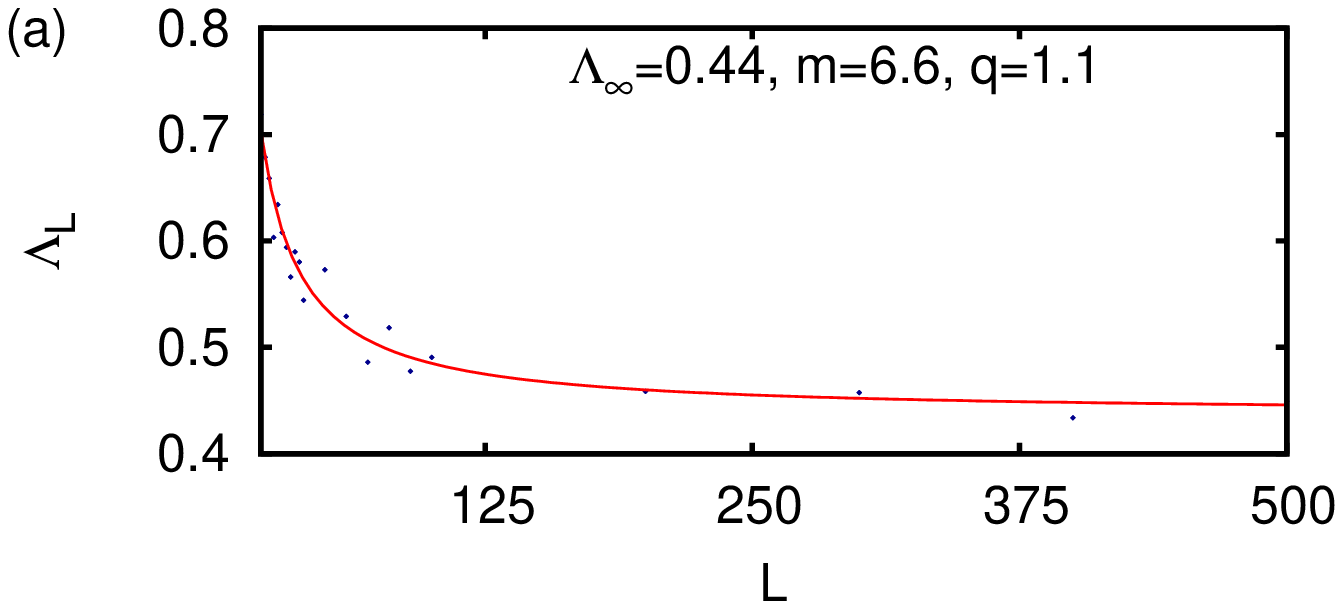}
\includegraphics[width=0.4\textwidth]{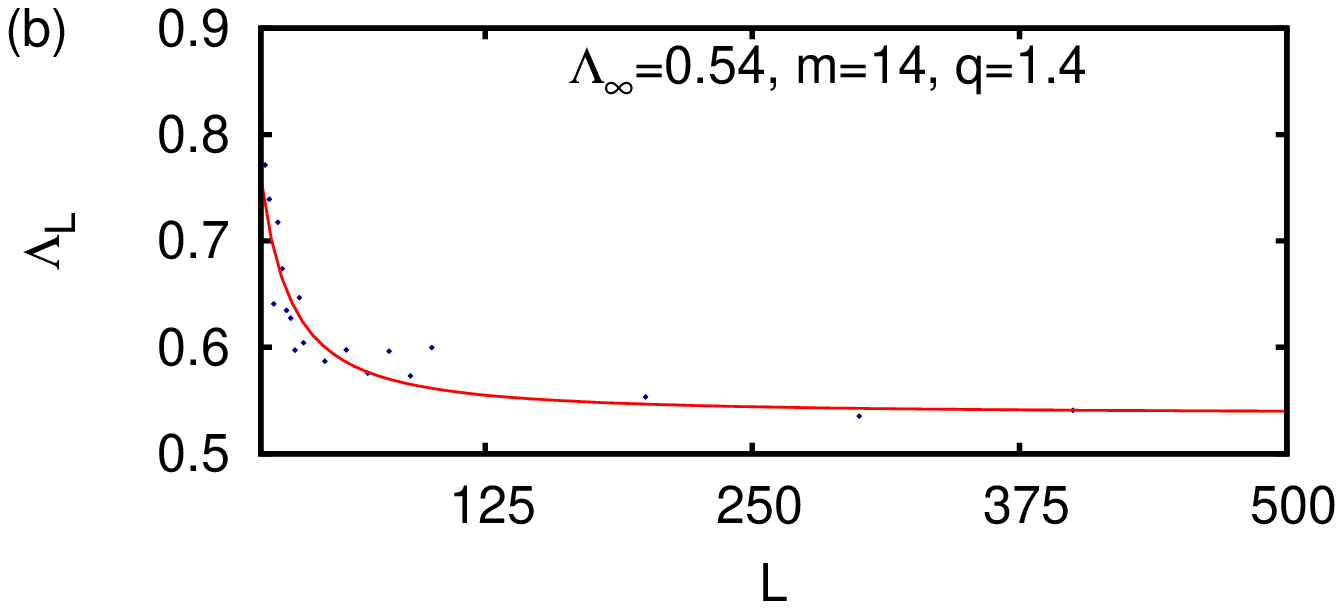}
\includegraphics[width=0.4\textwidth]{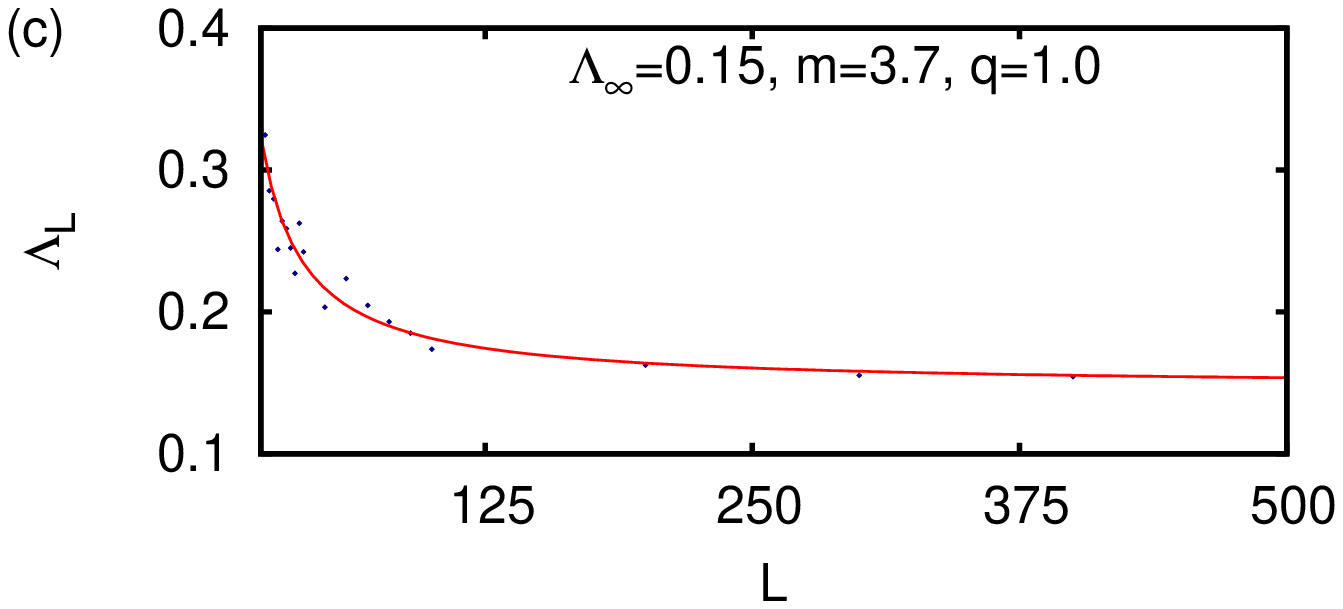}
\caption{$\Lambda_{L}$ obtained from the time series 
of $d_{1}(t)$ (blue crosses) and the fit 
$\Lambda_{\infty} + (m/L^{q})$ (red curve)  versus $L$, with 
$\omega_{0}= \omega_{1}=1$. 
Initial state $\ket{\alpha,1}\otimes\ket{\alpha}$ 
and (a) hopping frequency 
$\lambda=5$, $U=0.5$ (b) $\lambda=1$, $U=1$. 
 (c) Initial state $\ket{\alpha,5}\otimes\ket{\alpha,5}, 
 \lambda=1, U=1$.  $|\alpha|=1$ in all three cases.}
\label{fig:lambda_L_BEC}
\end{figure}

As in the foregoing,  we generate the 
time series of $d_{1}(t)$ by calculating this  difference 
for $20000$ time steps, in this case with $\delta t = 0.01$. 
As seen in Figs.  \ref{fig:lambda_L_BEC} (a)--(c), 
 in this instance $\Lambda_{L}$ is positive regardless of the degree of coherence of the initial states of the subsystems,  for a wide range of values of the ratio $U/\lambda_{1}$ 
(recall that  $\lambda_{1}= (\lambda^{2}+\omega_{1}^{2})^{1/2}$). 
 With an increase in $U/\lambda_{1}$,  
  $\Lambda_{\infty}$ increases (Figs. \ref{fig:lambda_L_BEC} (a), (b)). In contrast to the atom-field interaction model, a departure of the initial state from perfect coherence causes  $\Lambda_{\infty}$ 
  to decrease (Figs. \ref{fig:lambda_L_BEC} (a), (c)).

\begin{figure}[h]
\centering
\includegraphics[width=0.4\textwidth]{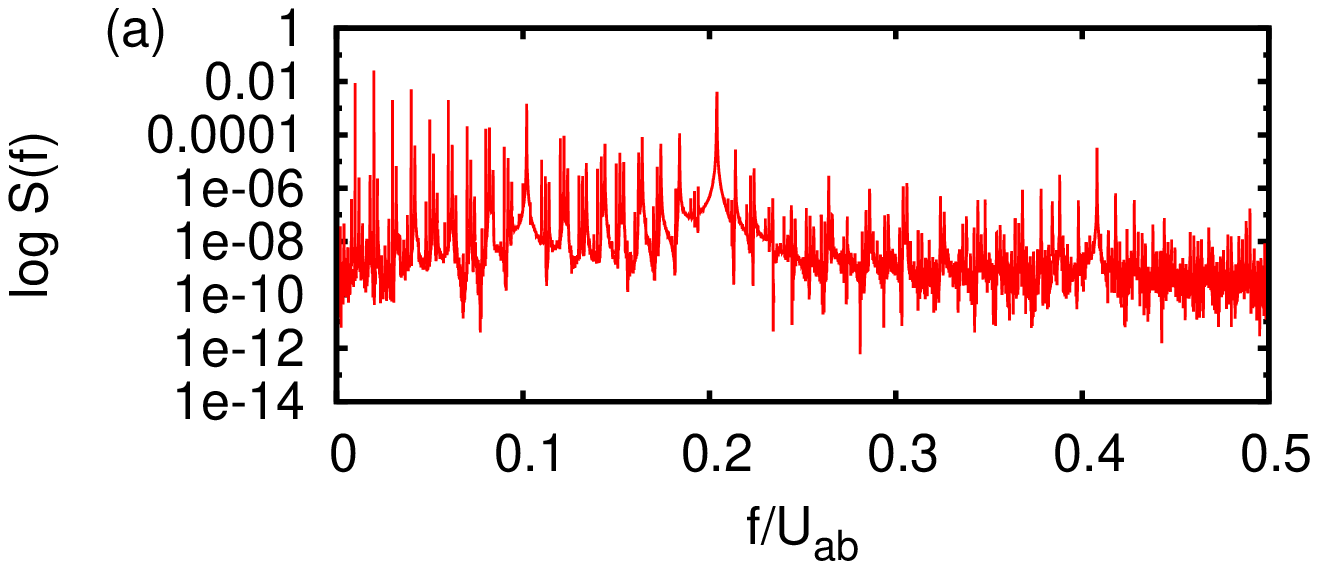}
\includegraphics[width=0.4\textwidth]{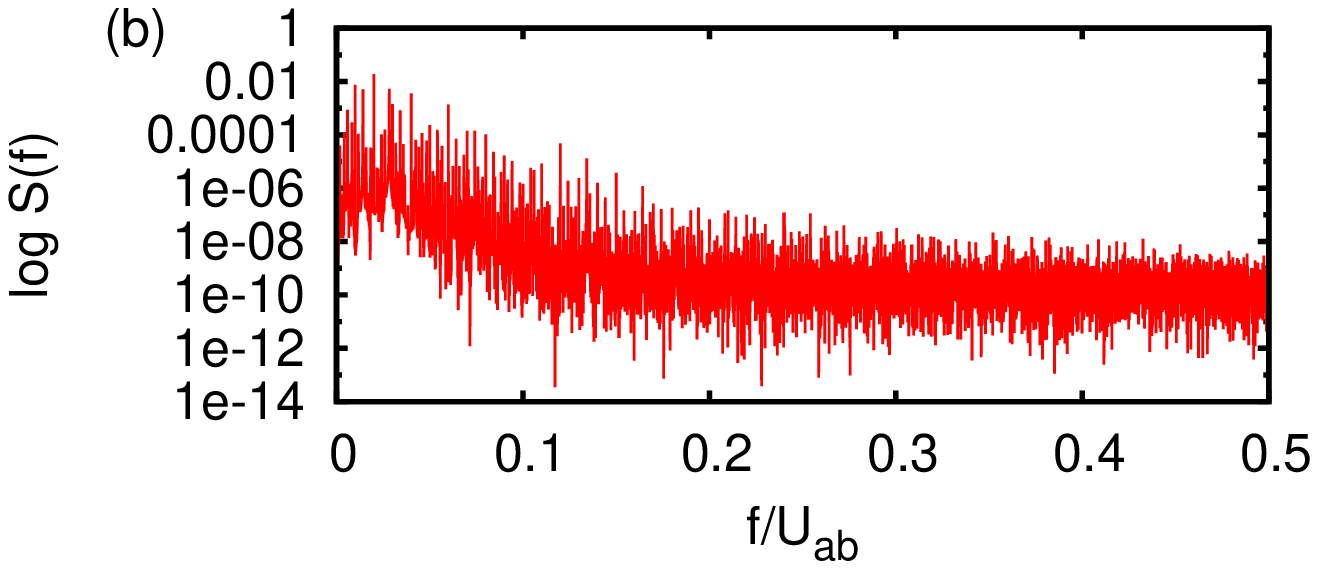}
\includegraphics[width=0.4\textwidth]{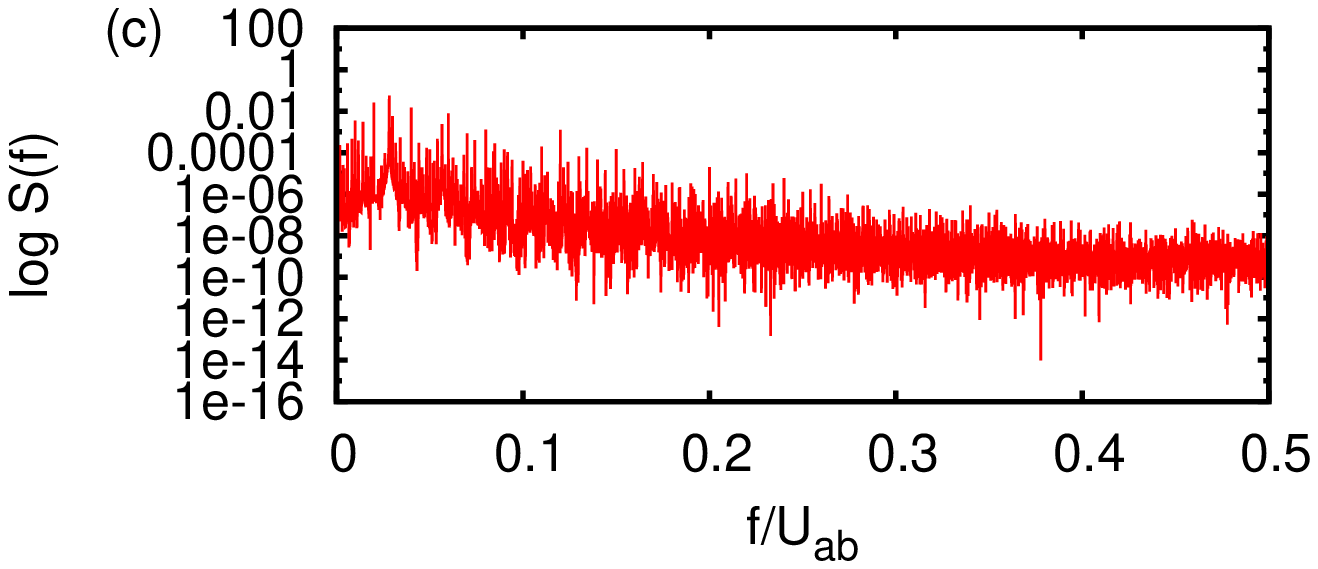}
\caption{Power spectrum $S(f)$ on a logarithmic 
 scale (red curve) versus  $f/U$ for the same 
 respective sets of parameters and initial states as in 
 Figs. \ref{fig:lambda_L_BEC} (a)--(c).}
\label{fig:powerspec_BEC}
\end{figure}

The power spectra  corresponding to  the three 
cases in Fig. \ref{fig:lambda_L_BEC} are shown in 
Fig. \ref{fig:powerspec_BEC}.  
When the linear 
part of $H_{\textsc{BEC}}$ is dominant  
($\lambda$ dominates over $U$, 
Fig. \ref{fig:powerspec_BEC} (a)), $S(f)$ reflects 
a degree of 
quasiperiodicity in the time series. When $U$ becomes 
comparable to $\lambda$, however, the nonlinearity 
in the Hamiltonian takes over, and $S(f)$ exhibits a 
broadband spectrum 
(Figs. \ref{fig:powerspec_BEC} (b), (c)).

\section{Concluding remarks}
\label{conclremarks}
We have investigated  
various features of an entanglement indicator for bipartite 
systems that is obtained \textit{solely} from tomograms. 
The  performance of  this indicator in quantifying the extent of entanglement has been assessed using two model bipartite systems with inherent nonlinearities. It has been shown that the indicator fares significantly better for generic initial states of the system even during temporal evolution, compared to better-known entanglement indicators such as one based on inverse participation ratios.  In order to quantify the reliability of the indicator over long intervals of time, the difference between the SVNE and our tomographic indicator has been examined using a time-series analysis. The manner in which this difference is sensitive to the nonlinearity of the system, the nature of the interaction, and the precise initial state is revealed by the time-series analysis. The importance and relevance of this investigation lies in the fact that detailed state reconstruction from the tomogram is completely avoided  in identifying an efficient entanglement indicator for generic bipartite systems involving continuous variables. 

\section*{Appendix: Time evolution in the double-well BEC model}
\label{sec:appen_1}

The double-well BEC effective Hamiltonian is given by Eq.  
(\ref{eqn:HBEC}). We require the state $\ket{\psi(t)}$ corresponding 
to an initial state that is a direct product of normalized 
boson-added coherent 
states of the atoms in the wells  $A$ and   $B$, namely,    
\begin{equation}
\ket{\psi_{m_{1},m_{2}}(0)}
 = \ket{\alpha_{a}, m_{1}} \otimes 
\ket{\alpha_{b}, m_{2}},
\label{}
\end{equation} 
where  $\alpha_{a}, \alpha_{b} \in \mathbb{C}$ and 
 $m_{1}, m_{2}$ are non-negative integers. The  dependence 
 of the state on $\alpha_{a}$ and $\alpha_{b}$ has been 
 suppressed on the 
 left-hand side for notational simplicity. The $m$-PACS 
 $\ket{\alpha, m}$ (defined in Eq. (\ref{mpacsdefn}))  
 reduces to the standard oscillator coherent state 
 $\ket{\alpha}$ for $m = 0$. 
 
 In the special case $m_{1} = 0, m_{2} = 0$, the state 
 at any time $t$ corresponding to the initial state 
 $\psi_{0,0}(0)$ has been shown in Ref. \cite{sanz} to be 
 given by
\begin{align}
\nonumber \ket{\psi_{00} (t)} =  
& e^{-\tfrac{1}{2}(|\alpha_{a}|^{2} + |\alpha_{b}|^{2})}  
\sum_{p,q=0}^{\infty} \frac{\beta_{1}^{p}(t) \,
\beta_{2}^{q}(t)}{\sqrt{p! q!}} \\
& e^{- i t (p+q)[\omega_{0} + U (p + q)]} 
\ket{p}\otimes\ket{q}, 
\label{eqn:Psi_t}
\end{align}
where 
\begin{equation}
\left.
\begin{array}{lll}
 \beta_{1}(t) & = & \alpha_{a} \cos \,(\lambda_{1} t) + 
(i/\lambda_{1}) ( \lambda \alpha_{b} - \omega_{1} \alpha_{a})
 \sin \,(\lambda_{1} t),\\[4pt] 
\beta_{2}(t) & = & \alpha_{b} \cos \,(\lambda_{1} t) + 
(i/\lambda_{1}) ( \lambda \alpha_{a} + \omega_{1} \alpha_{b})
 \sin \,(\lambda_{1} t),
 \end{array}
 \right\}
\label{beta2}
\end{equation}
and $\lambda_{1}=(\lambda^{2}+ \omega_{1}^{2})^{1/2}$. 
It can then be shown \cite{sharmila} that 
the state vector at time $t$ is given by 
\begin{equation}
\ket{\psi_{m_{1},m_{2}}(t)} 
= M_{m_{1},m_{2}}(t)\ket{\psi_{00}(t)}, 
\label{psit}
\end{equation}
where the operator $M_{m_{1},m_{2}}(t)$ is as follows.
Let $k, l, p, q$ denote non-negative integers, and let 
\begin{equation}
s = k+l+p+q, \,
\overline{p} =(k+m_{2}-l), \,\overline{q} =(l+m_{1}-k).  
\label{}
\end{equation}
Further, let 
\begin{equation}
\kappa= \big[m_{1}! m_{2}! L_{m_{1}}(-|\alpha_{a}|^{2}) 
 L_{m_{2}}(-|\alpha_{b}|^{2})\big]^{-1/2} 
\label{}
\end{equation}
and  $\Gamma=\cos^{-1}(\omega_{1}/\lambda_{1})$.
Then
\begin{eqnarray}
M_{m_{1},m_{2}}(t) & = & \kappa \Big\{\sum_{k=0}^{m_{1}}\sum_{l=0}^{m_{2}}\sum_{p=0}^{\overline{p}}\sum_{q=0}^{\overline{q}}(-1)^{k-p} 
{\tbinom{m_{1}}{k}} 
{\tbinom{m_{2}}{l}} \,
 {\tbinom{\overline{p}}{p}} \,
 {\tbinom{\overline{q}}{q}}\,
e^{2i(l-k)\lambda_{1} t}\times \nonumber \\[4pt]
&&(\cos\,\tfrac{1}{2}\Gamma)^{s} (\sin\,\tfrac{1}{2}\Gamma)^{2(m_{1}+m_{2})- s} (a^{\dagger})^{p+\overline{q}-q} \,(b^{\dagger})^{q+\overline{p}-p}\Big\}\times \nonumber\\[4pt] 
&&e^{-i\omega_{0} t(m_{1}+m_{2}) +i\lambda_{1}t (m_{1}- m_{2})}
e^{-i U t (m_{1}+m_{2}) (2 N_{\rm{tot}} + m_{1}+m_{2})}.
\label{eqn:intermed_rho_numerics}
\end{eqnarray}

\begin{acknowledgements}
One of the authors (SL) thanks M. Santhanam of IISER Pune for discussions pertaining to inverse participation ratios.
\end{acknowledgements}


%\bibliography{references}   % name your BibTeX data base


%\begin{thebibliography}{}
%\end{thebibliography}

\end{document}